\documentclass[a4paper,oneside,12pt]{amsart}
\usepackage{amsaddr} %
\usepackage{amsfonts}
\usepackage{amssymb}
\usepackage{amsmath}
\usepackage{amscd}
\usepackage{dsfont}
\usepackage{cite}
\usepackage[dvips]{graphicx}
\usepackage{color}

\newtheorem{cor}{Corollary}

\newtheorem{prop}[cor]{Proposition}

\newtheorem{defn}[]{Defenition}

\newtheorem{ex}[]{Example}

\newcommand{\norm}[1]{\left\Vert#1\right\Vert}

\newcommand{\set}[1]{\left\{#1\right\}}

\renewcommand{\i}{\textrm{i}}
\newcommand{\tr}{\textrm{Tr}\;}
\newcommand{\diag}[1]{\textrm{diag}\set{#1}}

\newcommand{\ben}[1]{\begin{equation}\label{#1}}
\newcommand{\een}{\end{equation}}
\newcommand{\be}{\begin{equation*}}
\newcommand{\ee}{\end{equation*}}
\newcommand{\bemn}[1]{\begin{multline}\label{#1}}
\newcommand{\eemn}{\end{multline}}
\newcommand{\bem}{\begin{multline*}}
\newcommand{\eem}{\end{multline*}}

\newcommand{\1}{\mathds{1}}
\newcommand{\p}{$p$}

\newcommand{\floor}[1]{\left\lfloor #1\right\rfloor}
\def\pclose{\stackrel{p}{\sim}}

\def\argmax{\mathrm{arg}\;\max}
\title[Wavelet analysis on sequences, two-fold de Bruijn sequences]{Wavelet analysis on symbolic sequences and two-fold de Bruijn sequences.}
\author[A]{V.Al. Osipov}
\address{Chemical Physics, Lund University, Getingev\"agen 60, 22241, Lund, Sweden}
\date {\today}

\begin{document}
\begin{abstract} The concept of symbolic sequences play important role in study of complex systems. In the work we are interested in ultrametric structure of the set of cyclic sequences naturally arising in theory of dynamical systems. Aimed at construction of analytic and numerical methods for investigation of clusters we introduce operator language on the space of symbolic sequences and propose an approach based on wavelet analysis for study of the cluster hierarchy. The analytic power of the approach is demonstrated by derivation of a formula for counting of {\it two-fold de Bruijn sequences}, the extension of the notion of de Bruijn sequences. Possible advantages of the developed description is also discussed in context of applied problem of construction of efficient DNA sequence assembly algorithms.
\end{abstract}


\maketitle
% ----------------------------------------------------------------
\setlength{\textfloatsep}{20pt plus 4pt minus 4pt}
\setlength{\floatsep}{20pt plus 4pt minus 4pt}
\setlength{\columnsep}{9pt}

\section{Motivation and structure.}
Nowadays symbolic sequences is a fundamental concept widely used in various fields of natural sciences~\cite{DFT2003, BH2007, B2000, K1998, HZ1998, H2009, B1946, BVKB2010, GR1961}. In bioinformatics,  theory of information, and theory of discreet Markov chains the intrinsic structure of objects under consideration provides direct mapping on symbolic sequences~\cite{K1998}. Less obvious extension of the symbolic approach one can find in study of complex behavior in dynamical systems~\cite{DFT2003, HZ1998}. In essence, the underlying idea is in organization of a stroboscopic sampling of the multidimensional trajectory. In case of Hamiltonian systems one constructs a Poincar\'e section surface in phase-space~\cite{T1989}, such that it is oriented orthogonal to the dynamical flow at each point. Linearization of the dynamics allows to separate stable and unstable directions of motion and thus define the set of feasible positions at the next crossing of the Poincar\'e section surface. All intersections falling within the same sub-region of the surface are designated by a certain symbol. For the system possessing a chaotic dynamics one can aver existence of the Markov partitioning of the surface, meaning that each next symbol in the symbolic dynamics is defined only by the previous one. For infinite or cyclic sequences the real trajectory can be uniquely restored from the symbolic dynamics. Often, the obtained Markov alphabet includes an infinite number of symbols, as, for instance, in chaotic billiards, out of the flow discontinuity on the billiard walls. It is not forbidden, however, to define some other partition prompted by physical (geometrical) properties of the system, which would generate a finite alphabet. The fee for such convenience is  formation of constraints on combinations of more then two subsequent symbols in the symbolic dynamics. In this article we consider a special case of symbolic dynamics generated by baker's map~\cite{O1993}. It is a chaotic map from the unit square into itself, its symbolic dynamics do not assume any constraints, i.e. all possible symbolic sequences correspond to some trajectory generated by the baker's map. 

\begin{figure}\centering
\includegraphics[scale=0.08]{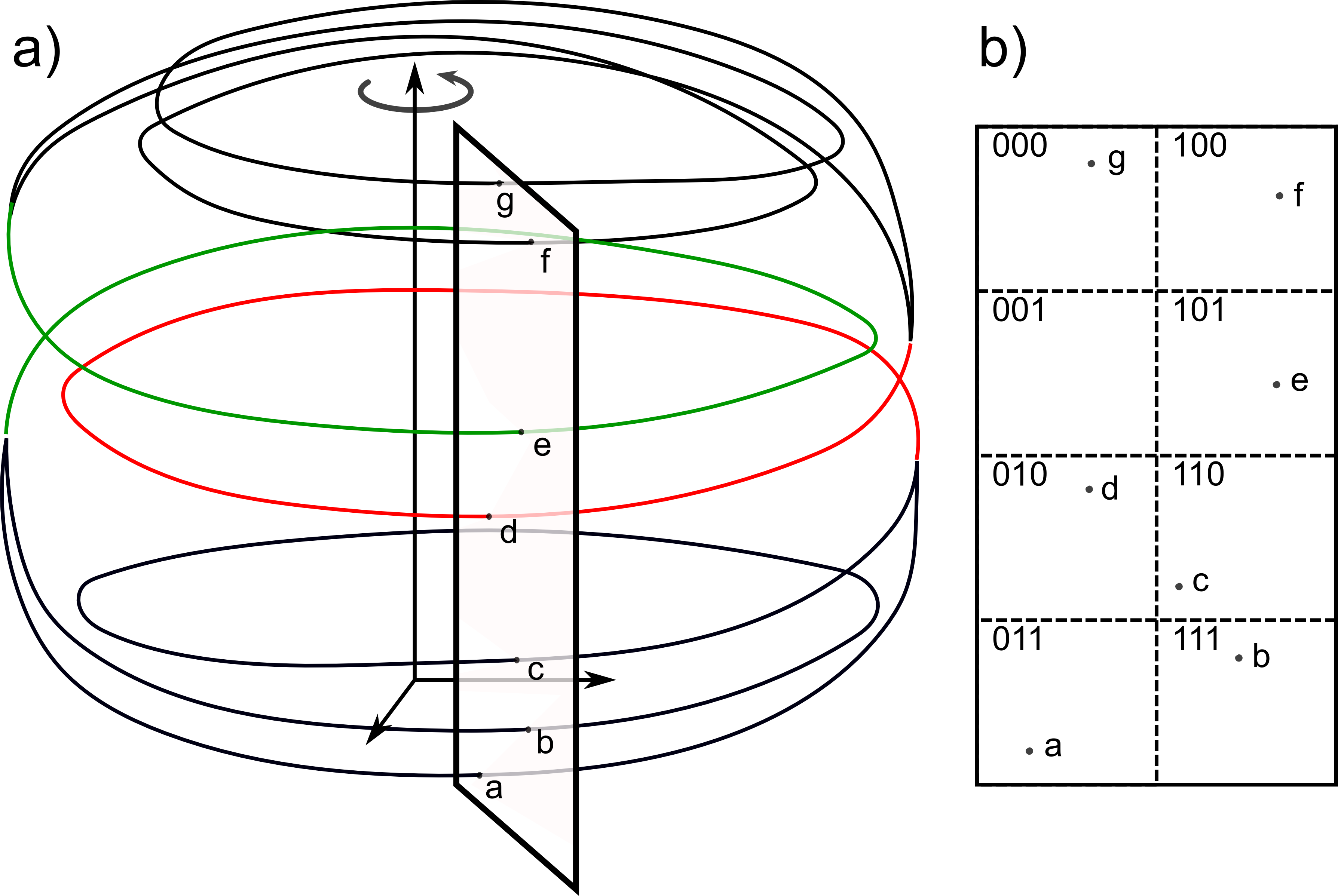}\hspace{20pt}\parbox[t]{3cm}{\includegraphics[scale=0.085]{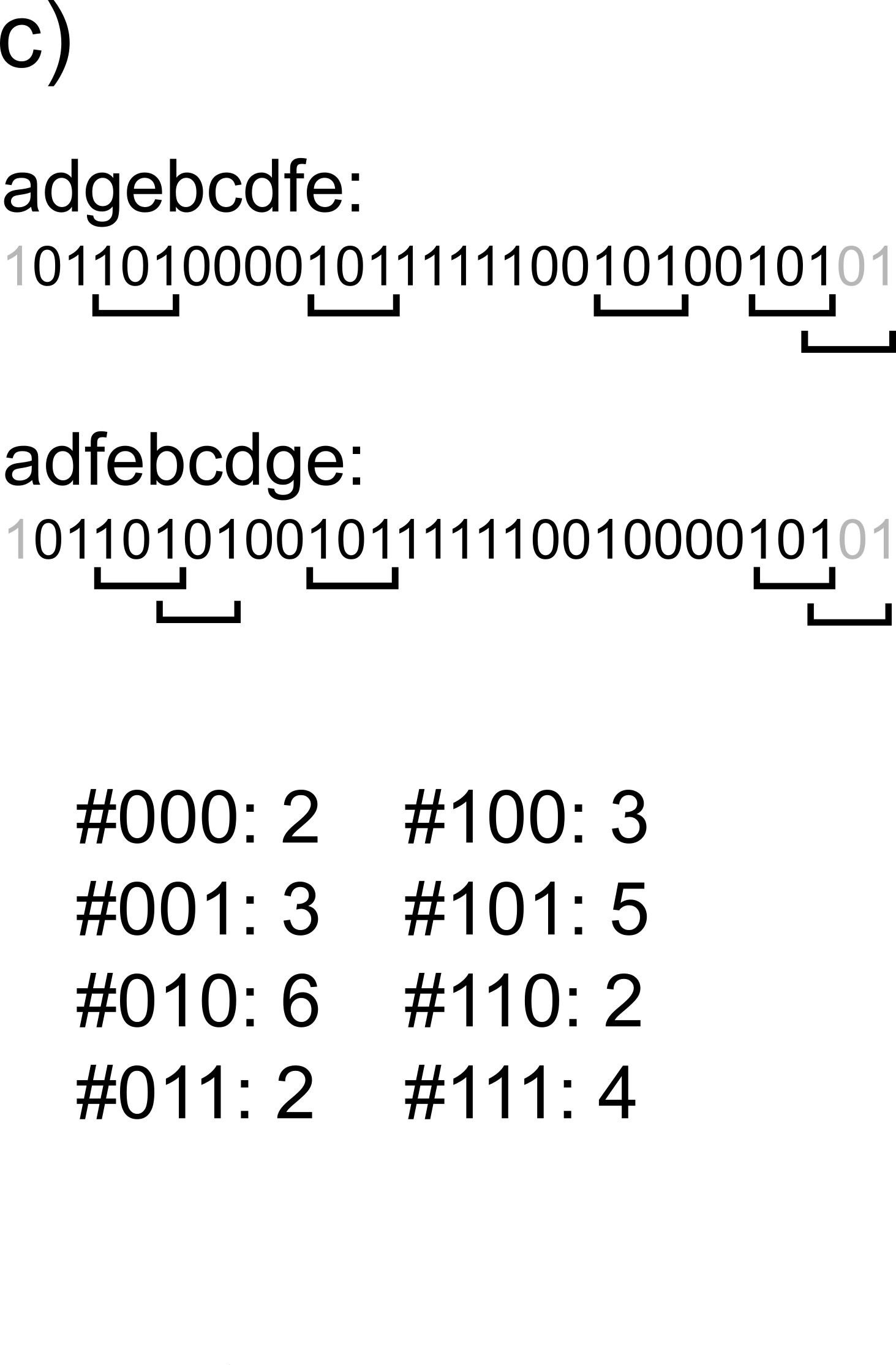}} \vspace{20pt}\\
\includegraphics[scale=0.25]{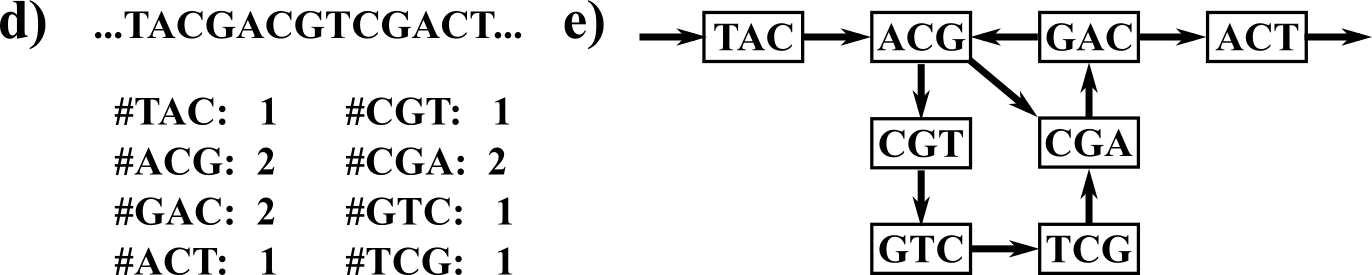}
\caption{\small \label{Fig4} a) Schematic representation of a periodic orbit with two encounters crossing the Poincar\'e section surface in points $e$ and $d$ (the colored loops). b) Possible choice of the finite partition on the Poincar\'e section surface. The 8 letters of the emerging alphabet are encoded by binary numbers. c) Symbolic representation for the pair of periodic orbits. The sequences are different but all triples of symbols containing in the upper sequence enter the same number of times into the lower one. d) A part of DNA sequence and the frequencies of the three-letters sub-strings. e) Graphical representation of the DNA sequences as a path on de Bruijn graph.}
\end{figure}

\begin{figure}\centering
\includegraphics[width=\textwidth]{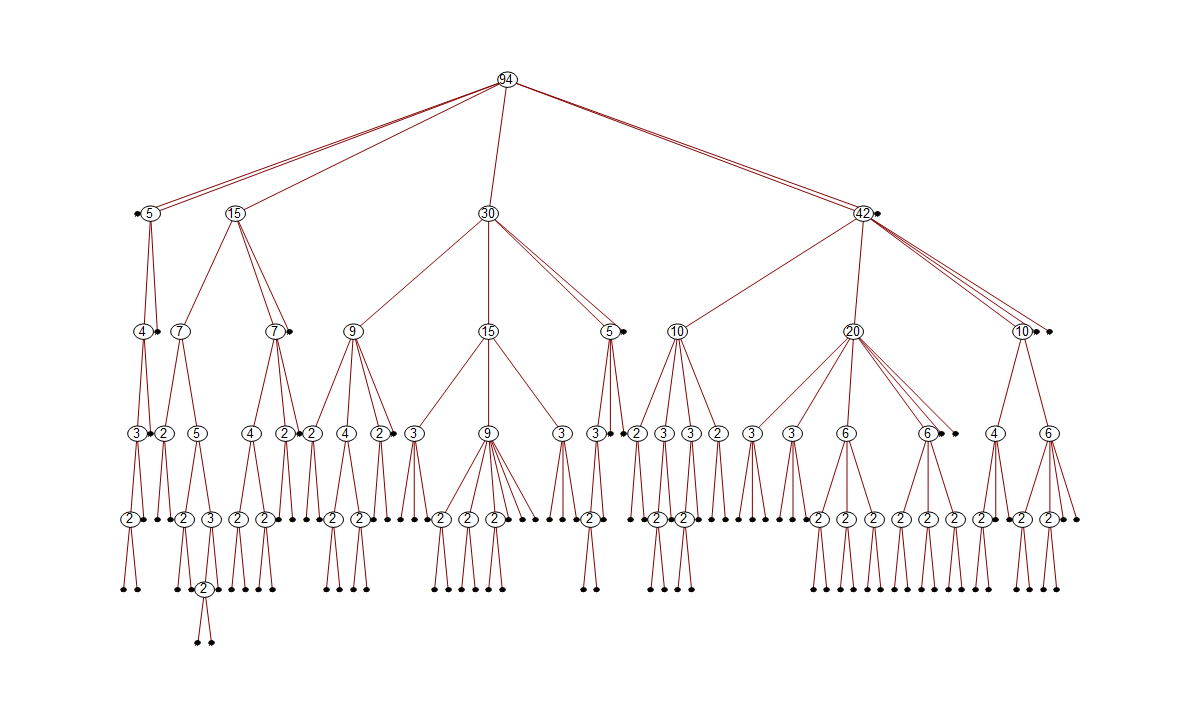}
\caption{\small\label{Fig5} Clustering of cyclic binary sequences of the length $n=11$ (only half of the tree, 94 sequences out of 188, is shown). The end-points of the tree (black circles) represent the sequences, the numbers in the circles give the total number of sequences in the corresponding cluster. The tree is constructed by the procedure developed in section 3. Cluster of sequences with $z=2$ ones and 9 zeros contains 5 sequences, with $z=3$ ones -- 15, with $z=4$ -- 30, with $z=5$ -- 42, the sequences with $z=0,1$ do not form non-trivial clusters, see eq.~(\ref{cluster}). Parameter $z$ determines the number of branches on the level $p=1$ of the tree (first from the top), the tree brunches up to the maximal level $p_{max}=5$, see section 3.4.}
\end{figure}

Study of periodic orbits and consequently {\it cycled} symbolic sequences plays a special role in theory of quantum chaotic systems~\cite{CAMTV2012}. Unstable periodic orbits form ``skeleton'' of a dynamical system and knowledge of their hierarchy can give one's access to many of the system's dynamical averages, such as, natural measure, Lyapunov exponents, fractal dimensions, entropy. They can efficiently expressed in terms of a sum over the unstable periodic orbits~\cite{A1987, A1990}. For instance, by means of semiclassical Gutzwiller trace formula~\cite{G1990} the eigenenergy density of chaotic quantum system can be expressed as a sum over long (of order Heisenberg time $\propto \hbar^{1-d}$, where $d$ dimension of the system) classical periodic orbits and, in turn, the density-density correlations as the double sum. Into the correlator formula the orbits come in pairs. The non-trivial oscillatory sub-leading terms in the correlator~\cite{H2010, SR2001, HMBH2004} result from interference between special type of periodic orbits, such that the partner orbits are close to each other everywhere in configuration space, but the same configuration points are visited in different time-order (see fig.~\ref{Fig4} a). The  time-order switching happens in the regions called encounters, where each of the orbits comes onto the shortest distance to itself, being, at the same time, mostly distanced from the partner. Symbolic representation of periodic orbits is not sensitive enough to distinguish points within the encounters and the free loops on the level of a single symbol  (fig.~\ref{Fig4} b). The closeness of the partner orbits is reflected in the fact that the corresponding symbolic sequences being globally different are locally identical, i.e. they both consist of the same set of sub-strings of some given length $p$ (fig.~\ref{Fig4} c). The above picture was recently formalized in~\cite{GO2013a} under the notion of \p-closeness: Two cyclic symbolic sequences $X$ and $Y$ of the same length $n$ are \p-close to each other ($X\pclose Y$) if any sub-string of the length $p$ appears both in $X$ and in $Y$ the same number of times (might be zero). The properties of the equivalence relation $\pclose$ allows to introduce an ultrametric distance $d(X,Y)$ on the set of periodic sequences. Such distance, in addition to the standard properties of distances, also satisfies the strong triangle inequality:  $d(X,Y)\le\max[d(X,Z),d(Z,Y)]$ (see for instance~\cite{M2008}). In other words, all cyclic sequences of a given length $n$ are distributed over hierarchically nested clusters with respect to their \p-closeness and thus they can be classified with respect to their local content, see fig.~\ref{Fig5}. The \p-closeness can be seen as generalization of de Bruijn sequences~\cite{B1946}, which first appearance can be traced back to the mathematical work of Flye Sainte-Marie~\cite{F1894} published in 1894. De Bruijn sequence is a cyclic sequence characterized by the property that each possible sub-string of a given length $p$ enters the sequence and only once. 

Interesting, that the similar need in searching of symbolic sequences from the set of their sub-strings emerges in area absolutely different from dynamical systems, the problem of contiguous DNA sequence assembly from the shorter DNA fragments~\cite{H2009}. In more detailed exposition, the procedure of genome reading represents a set of multiple readings of small DNA pieces called reads, while reconstruction of the whole original nucleotide sequence becomes an algorithmic problem of assembling a number of reads into a single sequence. In the modern {\it de novo} assembly approach the sequence assembly is considered as a search of an Eulerian cycle (path on a graph which traverses all edges and each only once) on a directed graph~\cite{CPT2011} (see fig.~\ref{Fig4} d,e), a sub-graph of de Bruijn graph (more precisely de Bruijn-Good graph~\cite{B1946a, G1946}). The vertices of the sub-graph are associated with the strings of a fixed length $p$ extracted from the set of all reads. Two vertices are connected if the corresponding strings share an exact $p-1$ overlap (exact definition of de Bruijn graph and its properties will be discussed in section 2.3). Taking into account that a certain sub-string can appear in the native DNA several times one have to deal with sequences possessing a certain encounter structure. There is a number of principal questions arising in the field, which has been reported in the literature: how the number of possible realizations of sequences behaves with the length of the sequences segments, $p$, and the related question what is the optimal length of the reads~\cite{CBP2009}? The computational bottleneck in realization of de novo algorithms is the amount of memory required for saving the de Bruijn sub-graph data, which, obviously, becomes central for assembling of large genomes~\cite{CB2011, PHCHTB2012, CLJSM2014}.

Generally speaking, both the study of topological structures of periodic orbits in dynamical systems and the problem of DNA sequence assembly are based on the same  mathematical apparatus. Our recent finding of the ultrametric structure of the set of symbolic sequences~\cite{GO2013a} allows to extend our understanding of the problems and develop novel analytic and algorithmic approaches. Some of these ideas we demonstrate in the present article. Moreover, we believe that the introduced below operator language and wavelet analysis on cyclic symbolic sequences will be useful as well for formulation of advanced theoretical approaches in theories describing multi-scale stochastic processes on the set of symbolic sequences (see for instance~\cite{MAL2005, AZ2007, AIMN2014, GO2011, DD2009}).

In section 2 we introduce the operator formalism for the work with symbolic sequences in their frequency representation and reproduce the main known facts about the ultrametric structure of sequences (section 2.2), their geometric representation by means of de Bruijn graphs, as well as formula for the number of de Bruijn sequences obtained by counting Euler and Hamiltonian cycles on the de Bruijn graphs (section 2.3). Here (section 2.4) we also repeat the known results regarding the distribution of clusters of symbolic sequences and derive some related basic formulae. The main analytic structures discussed in the present research are rising (section 2.2) and lowering (section 3.1) operators on the cluster tree. Analysis of them in basis of wavelets, constructed by analogy with $p$-adic wavelets~\cite{K2002, KKS2014} (see also~\cite{N2014}), allows to define rigorous procedure for reconstruction of sequences belonging to certain clusters (see example on fig.~\ref{Fig5}). According to the procedure, on each hierarchical level one reduces the problem to the solution of a number of independent linear systems of Diophantine equations (section 3.2). The main advantage of the approach is that after a certain hierarchy level $p$, such that $\ell^p>n$ ($\ell$ is the number of letters in alphabet, $n$ is the length of the sequence) there is only a finite number of relevant solutions and this number decreases as $p$ increases. As we show in section 3.3, the ``center of masses'' of our approach lies in analysis of connectivity of the obtained sequences. Such ``factorisation'' of the problem made possible derivation of a formula, eq.~(\ref{MainFormula}), for counting of two-fold binary de Bruijn sequences. The notion of two-fold de Bruijn sequences is introduced in section 3.5. Finally, in section~4 we discuss obtained results and consider possible applications of the revealed classification of sequences in various fields. 
  
\section{Set of cyclic symbolic sequences. Ultrametric structure of the set.}\label{Sec2}
\subsection{Primary definitions.}
Let $X_n^{ \mathcal{A}_\ell}$ denotes a set of all strings of $n$ characters chosen from $\ell$-symbols alphabet $\mathcal{A}_\ell$. The total number of such sequences is known to be $\ell^n$, we write $\norm{X_n^{\mathcal{A}_\ell}}=\ell^n$.  It is convenient to enumerate the letters of the alphabet $\mathcal{A}_\ell$ by integer numbers from zero to $\ell-1$. 
 
Let $V_{\ell^p}$ be an $\ell^p$-dimensional vector space with the scalar product defined in a standard way. Let the set $\set{e^{(p)}_1,\; e^{(p)}_2,\dots, e^{(p)}_{\ell^p}}$ be the elementary basis in the linear space $V_{\ell^p}$. Elementary basis consists of column-vectors, such that all entries of the $k$th vector, $e^{(p)}_k$, except the $k$th one, are zeros. Below, we also use notation $\set{e_k}_{k=1,\dots,\ell}$ for the elementary basis vectors of the space $V_\ell$.

Let us define the bijective transformation $\mathcal{V}$ between the symbolic sequences from $X_n^{ \mathcal{A}_\ell}$ and the basis vectors of the space $V_{\ell^p}$ by the rule
\begin{equation}\label{trivUltrametricity}
\mathcal{V}:\quad [\alpha_1,\alpha_2,\dots \alpha_n]\in X_n^{ \mathcal{A}_\ell}\to \bigotimes_{k=1}^n e_{1+\alpha_k}\equiv e^{(n)}_{1+\sum_{k=1}^n \alpha_k \ell^{n-k}}\in V_{\ell^n}.
\end{equation}
Here the symbol $\otimes$ stands for the Kronecker product, such that
$$
\left(\begin{array}{cc}
a&b\\
c&d
\end{array}\right)
\otimes
\left(\begin{array}{cc}
a'&b'\\
c'&d'
\end{array}\right)=\left(\begin{array}{cccc}
aa'&ab'&ba'&bb'\\
ac'&ad'&bc'&bd'\\
ca'&cb'&da'&db'\\
cc'&cd'&dc'&dd'\\
\end{array}\right).
$$
Representation of matrices in terms of Kronecker product is useful out of its basic properties: $(A\otimes B)\cdot(C\otimes D)=(A\cdot C)\otimes (B\cdot D)$ and $(A\otimes B)^\dag=A^\dag\otimes B^\dag$. Here and below the symbol $\dag$ stands for transponation and, if applicable, simultaneous complex conjugation.

To proceed with the formalization we introduce the reduction operator $\sigma_p: \quad V_{\ell^n} \to V_{\ell^p}$ for $p\le n$.  Contrary to the transformation $\mathcal{V}$ the aim of the latter construction is to project only the first $p$ symbols of the sequence $[\alpha_1,\alpha_2,\dots \alpha_n]\in X_n^{ \mathcal{A}_\ell}$ onto the vector space of reduced dimensionality, ignoring information of other symbols:
$$
\sigma_p\mathcal{V}  [\alpha_1,\alpha_2,\dots \alpha_n]= \bigotimes_{k=1}^p e_{1+\alpha_k}\equiv e^{(p)}_{1+\sum_{k=1}^p \alpha_k \ell^{p-k}}\in V_{\ell^p}.
$$
In terms of direct product the operator takes the following form (here and below $\1_k$ is $k\times k$ identity matrix and for matrices $\1_\ell$ we omit the index $\ell$)
$$
\sigma_p=\bigotimes_{k=1}^p \1\otimes \bigotimes_{k=p+1}^n q^\dag,
$$
where $q^\dag$ means the left multiplication on the vector of one's, i.e. $q^\dag=(e_1+e_2+\dots+ e_\ell)^\dag$.  It is assumed that at $p=n$ the matrix $\sigma_p$ coincides with the $\ell^n\times \ell^n$ identity matrix.

Let us finally introduce operator $T_n$ of a cyclic shift of symbols in symbolic sequence of the length $n$, such that:
$\mathcal{V}^{-1}T_n\mathcal{V} [\alpha_1,\alpha_2,\dots \alpha_n]\to [\alpha_2,\dots \alpha_n,\alpha_1]$. Its matrix representation in the basis $e^{(n)}_1$, ... $e^{(n)}_{\ell^n}$ is given by the formula
$$
T_n=\sum_{k=1}^\ell e_k \otimes \left( \bigotimes_{m=1}^{n-1}  \1\right) \otimes e^\dag_k.
$$

With the help of the shift operator one can define the set of cyclic sequences $\mathcal{X}_n^{ \mathcal{A}_\ell}$ as the factor set $X_n^{ \mathcal{A}_\ell}$ over $T_n$, $\mathcal{X}_n^{ \mathcal{A}_\ell}=X_n^{ \mathcal{A}_\ell}/T_n$ . The volume of the space can be estimated by the combinatorial formula known as the number of necklaces~\cite{St1999, GR1961} 
\begin{equation}\label{NumberSec}
\norm{\mathcal{X}_n^{ \mathcal{A}_\ell}}=\frac{1}{n}\sum_{d|n}\phi (d) \ell^{n/d}\qquad
\phi(d)=d\prod_{q|d}\left(1-\frac{1}{q}\right)
\end{equation}
where the sum runs over $d$, the divisors of $n$ (including $d=1$ and $d=n$), and $\phi(d)$ is the Euler's totient function, $q|d$ means that the product runs over all prime divisors of $d$. For $n$ -- prime the formula becomes particularly simple 
$$
\norm{\mathcal{X}_n^{ \mathcal{A}_\ell}} =\frac{\ell^n+(n-1)\ell}{n}\qquad n\mbox{ is prime.}
$$
Note also that from~(\ref{NumberSec}) one can see that the number of non-prime cyclic sequences is negligible in comparison with the number of prime sequences in the limit $n\to\infty$.

\subsection{Hierarchical structure of the set of cyclic symbolic sequences.}
In the article we are interested in the properties of the combined projecting operator 
\begin{equation}\label{Ppn}
P_{p,n}=\sigma_p  \sum_{k=0}^{n-1} T_n^k
\end{equation}
The operator $P_{p,n}$ acting on the vector $e^{(n)}_{1+\sum_{k=1}^n \alpha_k \ell^{n-k}}=\mathcal{V}[\alpha_1,\alpha_2,\dots,\alpha_n]\in V_{\ell^n}$ has the image in $V_{\ell^p}$. Its action on the basis vector $e^{(n)}_k$ produces a vector $x_{p,n}\in V_{\ell^p}$ with non-negative integer entries, we write $x_{p,n}\in\mathds{N}_0$. This set we denote as $\mathcal{F}_{p,n}$. The $j$th entry of the vector $x_{p,n}\in\mathcal{F}_{p,n}$ can be seen as frequency, i.e. the number of entrances (might be zero) of the string $[\beta_1,\dots,\beta_p]$ of the length $p$, such that $j=1+\sum_{k=1}^p \beta_k \ell^{p-k}$, into the original sequence $[\alpha_1,\alpha_2,\dots,\alpha_n]$. 

Consider some properties of $P_{p,n}$. Each vector $x_{p,n}\in\mathcal{F}_{p,n}$ has the property
\begin{equation}\label{xpnNorm}
\sum_{j=1}^{\ell^p}[x_{p,n}]_j=n.
\end{equation}
This follows from the fact that $P_{p,n}$ is a sum of $n$ terms (see eq.~(\ref{Ppn})), each of them produces an elementary basis vector in  $V_{\ell^p}$. In particular, action of $P_{p,n}$ on the vector corresponding to the sequence with identical symbols produces a vector proportional to $n$:
\begin{equation}\label{PpnNorm}
P_{p,n} e^{(n)}_{1+\alpha\sum_{k=1}^n \ell^{n-k}}=n e^{(p)}_{1+\alpha\sum_{k=1}^p \ell^{p-k}}\qquad \alpha=0,\dots,\ell-1.
\end{equation}
This is reflection of the generic property, that if some sequence ($e^{(n)}_k$) has the minimal period $d$, then all entries of $x_{p,n}=P_{p,n}e^{(n)}_k$ are proportional to $n/d$. 

All sequences obtained as a shift of a given one, i.e. vectors $e^{(n)}_k, T_ne^{(n)}_k,\dots, T^{n-1}e^{(n)}_k$, correspond to one and the same image $x_{p,n}=P_{p,n}e^{(n)}_k$. It follows directly from the structure of $P_{p,n}$. Below, to avoid this trivial degeneracy we prefer to work with the operator~(\ref{Ppn}) acting on the factor space $\mathcal{V}\mathcal{X}_n^{ \mathcal{A}_\ell}$. 

Let us now introduce a raising operator $R_{p-1}:\;\mathcal{F}_{p,n}\to \mathcal{F}_{p-1,n} $ which acts by the rule: for all $k=1,\dots, \ell^n$
\begin{equation}\label{raising}
\mbox{if }\qquad x_{p+1,n}=P_{p+1,n} e^{(n)}_k\qquad \mbox{and}\qquad x_{p,n}=P_{p,n} e^{(n)}_k,\qquad\mbox{  then }\qquad R_{p}x_{p+1,n}=x_{p,n}.
\end{equation}
Its matrix form can be read off from the relation $P_{p,n}=R_{p}P_{p+1,n}$, it has the form
\begin{equation}\label{raising2}
R_{p}=\bigotimes_{k=1}^{p} \1 \otimes q^\dag.
\end{equation}
Note, that the form of operator $R_p$ does not depend on $n$.

\begin{ex}
Let $n=3$, $\ell=2$. The number of sequences is $\norm{X_{3}^{(2)}}=8$, the size of the factor set $\norm{\mathcal{X}_{3}^{(2)}}=4$. The elementary basis has the form 
\begin{equation*}
e_1=\left(
\begin{matrix}
1\\
0
\end{matrix}
\right),\qquad e_2=\left(
\begin{matrix}
0\\
1
\end{matrix}
\right).
\end{equation*}
 The operators $T$ and $\sigma_2$ in this basis have the forms
\begin{equation*}
T=\left(
\begin{array}{cccccccc}
\mathbf 1 & 0 & 0 & 0 & 0 & 0 & 0 & 0 \\
 0 & 0 & \mathbf 1 & 0 & 0 & 0 & 0 & 0 \\
 0 & 0 & 0 & 0 & \mathbf 1 & 0 & 0 & 0 \\
 0 & 0 & 0 & 0 & 0 & 0 & \mathbf 1 & 0 \\
 0 & \mathbf 1 & 0 & 0 & 0 & 0 & 0 & 0 \\
 0 & 0 & 0 & \mathbf 1 & 0 & 0 & 0 & 0 \\
 0 & 0 & 0 & 0 & 0 & \mathbf 1 & 0 & 0 \\
 0 & 0 & 0 & 0 & 0 & 0 & 0 & \mathbf 1 \\
\end{array}
\right),\qquad \begin{array}{c}\sigma_1=\left(
\begin{array}{cccccccc}
  \mathbf 1 &  \mathbf 1 &  \mathbf 1 &  \mathbf 1 & 0 & 0 & 0 & 0 \\
 0 & 0 & 0 & 0 & \mathbf 1 & \mathbf 1 & \mathbf 1 & \mathbf 1 \\
\end{array}
\right)\\\\
\sigma_2=\left(
\begin{array}{cccccccc}
  \mathbf 1 & \mathbf 1 & 0 & 0 & 0 & 0 & 0 & 0 \\
 0 & 0 & \mathbf 1 & \mathbf 1 & 0 & 0 & 0 & 0 \\
 0 & 0 & 0 & 0 & \mathbf 1 & \mathbf 1 & 0 & 0 \\
 0 & 0 & 0 & 0 & 0 & 0 & \mathbf 1 & \mathbf 1 \\
\end{array}
\right)
\end{array}
\end{equation*}
 Operator $P_p$ has the following matrix representation if it acts in the space $\mathcal{V} X_{3}^{(2)}$ (on the left) and in the space $\mathcal{V} \mathcal{X}_{3}^{(2)}$ (the right column)
\begin{equation*}
\begin{array}{c}P_{1,3}=\left(
\begin{array}{cccccccc}
 3 & 2 & 2 & 1 & 2 & 1 & 1 & 0 \\
 0 & 1 & 1 & 2 & 1 & 2 & 2 & 3 \\
\end{array}
\right)\\\\
P_{2,3}=\left(
\begin{array}{cccccccc}
 3 & 1 & 1 & 0 & 1 & 0 & 0 & 0 \\
 0 & 1 & 1 & 1 & 1 & 1 & 1 & 0 \\
 0 & 1 & 1 & 1 & 1 & 1 & 1 & 0 \\
 0 & 0 & 0 & 1 & 0 & 1 & 1 & 3 \\
\end{array}
\right)
\end{array}\qquad
\begin{array}{c} P_{1,3}=\left(
\begin{array}{cccc}
 3 & 2 & 1 & 0 \\
 0 & 1 & 2 & 3 \\
\end{array}
\right)\\\\
 P_{2,3}=\left(
\begin{array}{cccc}
 3 & 1 & 0 & 0 \\
 0 & 1 & 1 & 0 \\
 0 & 1 & 1 & 0 \\
 0 & 0 & 1 & 3 \\
\end{array}
\right)
\end{array}
\end{equation*}
As one can see vectors from the sets $\mathcal{F}_{1,3}$ and $\mathcal{F}_{2,3}$ have integer entries and support the general results~(\ref{xpnNorm}) and~(\ref{PpnNorm}). One can reveal (generically not uniquely) the corresponding symbolic sequence form $x_{p,n}$. For instance,
$$
x_{2,3}=(1\;1\;1\;0):\qquad \begin{array}{cccccc}
j=1+\sum_{k=1}^2 \beta_k 2^{2-k}&&\beta_1&\beta_2&&[x_{2,3}]_j\\
1&&0&0&&1\\
2&&0&1&&1\\
3&&1&0&&1\\
4&&1&1&&0
\end{array}
$$
correspond to the cyclic sequence $[001]$.
\end{ex}

Now we are at position to discuss the hierarchical structure of the set of cyclic sequences. First we introduce the notion of \p-closeness (see~\cite{GO2013a}).

\begin{defn}\label{pclose} Two sequences $a,b \in\mathcal{X}_n^{\mathcal{A}_\ell}$ are \p-close, $a\pclose b$, if their images $x_{p,n}=P_{p,n} \mathcal{V} a$ and $y_{p,n}=P_{p,n}  \mathcal{V} b$ are equal $x_{p,n}= y_{p,n}$.
\end{defn}

There are three important properties of the equivalence relation $\pclose$:
\begin{itemize}
\item[i.]
The relations $a \pclose b$ and  $a  \pclose c$ also imply that  $b \pclose c$;
\item[ii.]
The relation  $a \pclose b$ implies that $a \stackrel{p-1}{\sim} b$.
\item[iii.]
The relation  $a \stackrel{p-1}{\sim} b$ {\it does not} imply that $a\pclose b$.
\end{itemize}
The first is the direct sequence of the definition~\ref{pclose}, while the second follows from the linear properties of the  raising operator~(\ref{raising}),~(\ref{raising2}). The third sentence, in particular, means that the lowering operator translating $x_{p,n}=P_{p,n} e^{(n)}_k$ into $x_{p+1,n}=P_{p+1,n} e^{(n)}_k$ cannot be constructed as a linear operator. The latter claim will be clarified in the part 3.

From the properties i, ii, iii it follows that all cyclic sequences can be distributed over hierarchically nested clusters with respect to their closeness. Using the notion of \p-closeness one can naturally introduce the ultrametric distance on the set $\mathcal{X}_n^{\mathcal{A}_\ell}$ in the following way: The ultrametric distance $d(a,b)$ between  $a,\;b\in \mathcal{X}_{n}^{ \mathcal{A}_\ell}$ is defined as
\begin{equation}\label{ultrdist}
d(a,b)=e^{-\gamma_{max}(a,b)}\quad \gamma_{max}(a,b)=\argmax_{p=0,\dots,n-1}\set{p: a\pclose b}\quad \mbox{and}\quad d(a,b)=0\quad \mbox{if}\quad a=b. 
\end{equation}
The distance $d(a,b)$ is positive, symmetric and satisfies the strong triangle inequality:
$$
d(a,b)\le \max\set{d(a,c),\;d(b,c)}
$$
The latter can be easily proved. Assume that $a\pclose c$, $b \stackrel{p'}{\sim}c$ but $b \stackrel{p'+1}{\not\sim}c$ and $p'>p$. From the definition of \p-closeness it follows that $a\stackrel{p'}{\sim}c$ and therefore $a\stackrel{p'}{\sim}b$ but $a \stackrel{p'+1}{\not\sim}b$.

The function $\gamma_{max}$ can be calculated with the help of the operator $R_p$. If $x_{n,n},\;x'_{n,n}\in\mathcal{F}_{n,n}$ are respectively represent the sequences $a$ and $b$ then 
$$
\gamma_{max}=\argmax_{p=0,\dots,n-1}\set{p:\;R_p\cdot R_{p+1}\dots R_{n-1} (x_{n,n}- x'_{n,n})=0}.
$$

Before we start our study of the clusters it is instructive to discuss first the graphical representation of sequences and some basic results.

\subsection{Geometric representation of the cyclic sequences. De Bruijn sequences.}
The simplest way to represent the set of sequences $X_n^{ \mathcal{A}_\ell}$ graphically is to use regularly branching tree. Setting the ``left'' end of the sequence onto the root (the top point) of the tree one goes down choosing the branches according to the current letter within the sequence (see appendix~\ref{AppTree}). Thus the ``leaves'' are enumerated by integer numbers from $1$ to $\ell^n$, i.e. by the indexes of the elementary basis vectors of the space $V_{\ell^n}$. The obtained tree obviously generates ultrametrics on the set of symbolic sequences, {\it but} does not reflect the idea of equivalence of symbols with respect to their positions in the sequence and dissimilarity of symbols with respect to their local surrounding. To this end a more adequate graphical structure, de Bruijn graph, has been proposed~\cite{B1946a}. 

\begin{figure}
\centering \includegraphics[scale=0.5]{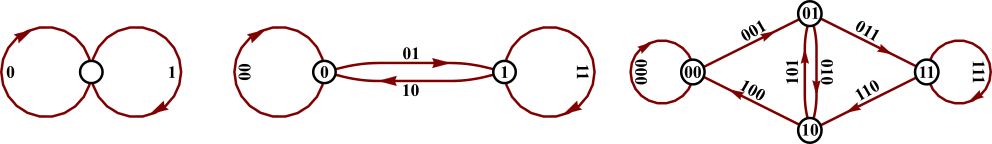}
\caption{\small \label{Fig6} De Bruijn graphs, from left to right, $G_2(0)$, $G_2(1)$, $G_2(3)$. The edges and vertices are encoded by binary sequences, see text for explanation.}
\end{figure}

The notion of de Bruijn graph is used here for a class of directed graphs $G_\ell(p)$ labeled by the integer numbers $p$ and $\ell$. Graph $G_\ell(p)$ has $\ell^p$ vertices connected by $\ell^{p+1}$ directed edges. Each vertex is designated with the symbolic string from $X_p^{ \mathcal{A}_\ell}$, figure~\ref{Fig6}. For our purposes, it is convenient also to mark the vertices by the corresponding basis vectors $e^{(p)}_k\in V_{\ell^p}$ (exceptions are the graphs $G_\ell(0)$ and $G_\ell(1)$, otherwise one has to work with zero- and one-dimensional spaces), where the vector index $k$ is calculated according to the rule~(\ref{trivUltrametricity}). Two vertices $e^{(p)}_k$ and $e^{(p)}_{k'}$ are connected by the edge with the label $e^{(p+1)}_{\ell k +j-1}=e^{(p)}_k\otimes e_j\in V_{\ell^{p+1}}$ ($j=1,\dots,\ell$) iff $e^{(p)}_{k'}=L_p e^{(p)}_k\otimes e_j$, where $L_p$ is the incidence matrix:
$$
L_p=q^\dag\otimes\left(\bigotimes_{k=1}^p \1\right).
$$  

Consider now some properties of the graph $G_\ell(p)$.  First, vertices have exactly $\ell$ outgoing and $\ell$ incoming edges. Therefore, the graph is Eulerian, i.e. one can find a path through all vertices, such that it traverses each edge only once.  Second, the $\ell^p\times \ell^p$ adjacency matrix $Q_{\ell^p}$ of the graph $G_\ell(p)$ can be restored from matrices $R_p$ and $L_p$. Indeed, starting from some vertex $e^{(p)}_k$ we obtain all possible outgoing edges by the formula $T_{p+1}\cdot q\otimes e^{(p)}_k=R^\dag_{p} e^{(p)}_k$. The backward projection onto the set of vertices can be done with the help of $L_p$.  Finally we obtain
\begin{equation}\label{Qellp}
Q_{\ell^p}=q^\dag\otimes \left(\bigotimes_{k=1}^{p-1} \1\right)\otimes q=L_pR^\dag_p.
\end{equation}
By analogy one can derive the expression for the connectivity $\ell^{p+1}\times \ell^{p+1}$ matrix $\tilde Q_{\ell^p}$ (the adjacency matrix for edges). It is $\tilde Q_{\ell^p}=T_{p+2}\cdot q\otimes L_p=L_{p+1}R^\dag_{p+1}$, which is,  on the other hand, the vertex adjacency matrix $Q_{\ell^{p+1}}$ of the graph $G_\ell(p+1)$. Due to equivalence $\tilde Q_{\ell^p}=Q_{\ell^{p+1}}$ via the line graph construction~\cite{W} each graph $G_\ell(p)$ is Eulerian and Hamiltonian. The number of Hamiltonian paths on the graph $G_\ell(p+1)$ is equal to the number of Eulerian paths on the graph $G_\ell(p)$.

The sequences corresponding to the Eulerian paths on de Bruijn graph  have a joint name of de Bruijn sequences~\cite{B1946}. The de Bruijn sequence is characterized by the fact that each possible sub-string of the length $p$ enters the sequence and only once, therefore its length is $n=\ell^p$. Below we give definition of de Bruijn sequences in terms introduced in this article and as a special case of $f$-fold de Bruijn sequences.
\begin{defn}
For given positive integer $p$, and $f$, the $p$-ary $f$-fold de Bruijn sequences from the set $\mathcal{X}_{f\ell^p}^{\mathcal{A}_\ell}$ is the set of \p-close sequences corresponding to the frequency vector
$$x_{f\ell^p,p}=f\cdot\bigotimes_{k=1}^p q;$$
the set of $p$-ary single-fold de Bruijn sequences coincide with the set of $p$-ary de Bruijn sequences.  
\end{defn}

The number of $p$-ary single-fold de Bruijn sequences can be calculated with the help of BEST theorem~\cite{AEB1951,St1999}. It states that the number of Eulerian cycles for some directed graph $G(v_i)$ with $n$ vertices $v_i$ can be calculated by the formula:
\begin{equation}\label{EulerCycles}
Co_k[D-Adj(v_i)]\prod_{i=1}^n(d_{out}(v_i)-1)!\, ,
\end{equation}
where $Adj(v_i)$ is the adjacency matrix, $D=\diag{\sum_{j=1}^n \big[Adj(v_i)\big]_{i,j}}_{i=1,n}$, the factor $d_{out}(v_i)$ is the number of outgoing edges at vertex $v_i$. The abbreviation $Co$ stands for cofactor taken for some arbitrary chosen raw $i$. In case of  $G_\ell(p-1)$ the second multiplier gives $(\ell!)^{\ell^{p-1}}/\ell^{\ell^{p-1}}$. Calculation of the cofactor means that one should remove from $Q_p$ the elements in the $k$-th raw and $k$-th column. Let $k=1$, then one obtains
\begin{equation}\label{cofactor}
Co_1[D-Q_{\ell^{p-1}}]=\det\left[\ell \bigotimes_{k=1}^{p-1} \1-Q_{\ell^{p-1}}^*\right],
\end{equation}
where the matrix $Q_{\ell^{p-1}}^*$ is obtained from $Q_{\ell^{p-1}}$ by dropping the first raw and the first line. The structure of the graph manifests itself in the special relations for traces, $\tr ( Q_{\ell^{p-1}}^*)^m=\ell^m-1$, $m=1,\dots,p-1$, and in a special form of the secular equation~\cite{R2002} for  $Q_{\ell^{p-1}}^*$:
$$
\lambda^{\ell^{p-1}-p-3}\left(\lambda^{p}-(\ell-1)\sum_{k=1}^{p-1}\lambda^k\right)=0.
$$
Substitution of $\lambda=\ell$ into the left-hand-side of the above equation allows us to calculate the sought cofactor and to get the famous formula  for the number of the $p$-ary de Bruijn sequences, it is  $\frac{(\ell !)^{\ell^{p-1}}}{\ell^{p}}$. First the formula was obtained for $\ell=2$ in~\cite{F1894} and later extended for generic $\ell$~\cite{B1946}.

\subsection{Clusters of cyclic symbolic sequences.}
Here we reproduce some results describing the distribution of clusters, see particular realization of the cluster tree on the fig.~\ref{Fig5}. 

The zero level ($p=0$) of the cluster tree contains only one cluster $\mathcal{C}_0$ of all cyclic sequences of the length $n$. The number of such sequences is given by the formula~(\ref{NumberSec}). The first level ($p=1$) consists of  the clusters, which number is given by the simple combinatorial formula $\frac{(n+\ell-1)!}{(\ell-1)! n!}$. The number of sequences in each cluster of the level $p=1$ can be calculated by Burnside's lemma~\cite{St1999}, it extends the result~(\ref{NumberSec}). Let the frequency vector $x_{n,1}$ is equal to $[a_1,a_2,\dots, a_\ell]$, then the number of sequences can be calculated from the expression
\begin{equation}\label{cluster}
\norm{\mathcal{C}\big[x_{n,1}=[a_1,a_2,\dots, a_\ell]\big]}=\frac{1}{n}\sum_{d|\set{a_1,a_2,\dots, a_\ell}}\phi (d) \frac{(n/d)!}{(a_1/d)!\dots(a_\ell/d)!},
\end{equation}
where $d$ are common divisors of $a_1,a_2,\dots, a_\ell$. Since $\sum_{d|n} \phi (d)=n$ we come to the obvious conclusion, that the cluster corresponding to the sequence of identical elements contains only one sequence. The same result one obtains for the sequence with only one non-identical element ($\phi (1)=1$).

There are some known asymptotic results obtained for the distribution of clusters in case of two-letters alphabet~\cite{GO2013a, GO2013b}. In particular, by means of heuristic arguments it was shown that the total number of clusters in the regime of finite $p$ and large $n$ grows  as $w_p n^{2^{p-1}}(1+O(1/n))$, where the coefficient ${w}_p$ can be calculated explicitly for small values of $p$ ($w_2=1/4$). The asymptotics for the average sizes of clusters, $<\norm{\mathcal{C}}>$, at different ratios of $p$ to $n$ were derived with the help of the random matrix theory methods. The asymptotic of $<\norm{\mathcal{C}}>$ solely depends on the average number of encounters. For $n \lesssim \sqrt{2^p}$, when the encounters are rare, i.e. the majority of frequency vectors from $\mathcal{F}_{p,n}$ are {$(0,1)$-vectors}, the average cluster size is one, {$<\norm{\mathcal{C}}>\approx 1$}, and the number of clusters is almost equal to the total number of cyclic sequences. Decreasing of the parameter $p$ leads to formation of encounters. 
As the average number of encounters becomes of the order one, an exponential growth of cluster sizes starts. For $2^{p/2}\ll n\ll 2^p$  one can find that $\log<\norm{\mathcal{C}}> \sim n^2$. In the regime $n\gg 2^p$ the average size is $<\norm{\mathcal{C}}>=\frac{2^{n}}{n}\left(\frac{2^{p-1}}{n\pi}\right)^{2^{p-2}}  (1+O(1/n))$. In this regime almost every point of a generic cyclic sequence belongs to some encounter and the further growth of $n$ does not lead to any essential increase of the number of encounters. This, in turn, results in a slower growth of cluster sizes, $\log<\norm{\mathcal{C}}> \sim n$. 

The size of the largest cluster, $\norm{\mathcal{C}_{\max}}$ in the limit of long sequences $n\gg1$ and finite $p$ asymptotically behaves like   $\norm{\mathcal{C}_{\max}}= \left(\frac{2^{n}}{n}\right)\left(\frac{2^p}{\pi n}\right)^{2^{p-2}}\left(1+O(n^{-1})\right)$. Note, that to the leading order of $n$ the periodic orbits belonging to the largest cluster cover the graph $G_2(p)$ uniformly, i.e. the elements of the corresponding frequency vector $x_{p,n}$ are almost identical. The latter observation stimulated our interest to the calculation of the number of two-fold de Bruijn sequences (see section~3.4).

Generically the distribution of the cluster sizes in the limit of long sequences and finite $p$ can be described by the
probability density that a randomly chosen cyclic sequence belongs to a cluster of the size less then $t \norm{\mathcal{C}_{\max}}$, $t\in [0,1]$, it is
$$
\rho(t)=\frac{\left(-\log t\right)^{2^{p-2}-1}}{(2^{p-2}-1)!}.
$$
One can also derive the expression for the probability of finding $k$ randomly chosen sequences of the same length to be belonging to the same cluster, 
$$
 \left( \frac{1}{k}\right)^{2^{p-2}}\left(\frac{2^p}{\pi n}\right)^{(k-1)2^{p-2}}\left(1+O(n^{-1})\right).
$$
The above two results were derived from estimation of the moments of cluster sizes in~\cite{GO2013a}. The second moment was previously considered in~\cite{S2010, T2000}. Classification of long periodic orbits with respect to the number of their encounters and approaches for counting of the corresponding cluster sizes one can find in~\cite{NBMSHH2007}. For instance, the clusters of sequences having $2m$ encounters well separated by non-intersecting free parts (see fig.~\ref{Fig4}) contain $\frac{(2m)!}{2^{2m}(2m+1)!}$ sequences. The formula for the number of short length binary cyclic sequences ($n\sim p$) has been derived in~\cite{WXY1986}.

\section{The lowering operator. Two-fold de Bruijn sequences.}\label{Sec3}
\subsection{The lowering operator $\mathcal{F}_{p,n}\to \mathcal{F}_{p+1,n}$.}
It is clear that the lowering operator cannot be organized by the same simple way as $R_{p}$. It is a non-linear multi-valued operator, generating a set of vectors $x_{p+1,n}\in\mathcal{F}_{p+1,n}$ out of the single vector $x_{p,n}\in\mathcal{F}_{p,n}$. Thus the problem of finding the lowering operator is, in fact, the problem of restoring information from the data given in integrated form. To this end, we propose an algorithmic solution. 

To simplify notations it is assumed along this section that $n$ is fixed and we settle the following notations
\begin{itemize}
\item[] Let $X\equiv x_{p-1,n}\in\mathcal{F}_{p-1,n}$;
\item[] Let $Y\equiv x_{p,n}\in\mathcal{F}_{p,n}$, such that $R_{p-1}Y=X$;
\item[] Let $\set{\bar{Z}}\subset V_{\ell^{p+1}}\cap \mathds{N}_0$, such that for all $\bar Z$, $R_{p}\bar Z=Y$;
\item[] Let $\mathcal{F}_{p+1}[Y]\equiv\set{\bar Z}\cap \mathcal{F}_{p+1,n}$.
\end{itemize}

The lowering operator acts in two steps  
\begin{itemize}
\item[I.] Restoring of the set $\set{\bar Z}$ from the information integrated in  $Y$;
\item[II.] Taking the operation of intersection in $\set{\bar Z}\cap \mathcal{F}_{p+1,n}$ to obtain $\mathcal{F}_{p+1}[Y]$.
\end{itemize}
Consider both steps separately.

\subsection{The first step of the algorithm.}Here we formulate a set of equations which generate all vectors from $\set{\bar Z}$ and demonstrate the most moderate method to resolve them based on the usage of wavelet basis.

\begin{prop}
Elements of the set $\set{\bar Z}$ satisfy the following set of Diophantine equations
\begin{eqnarray}\label{eq1}
L_p \bar Z&=&Y;\\\label{initialization}
R_p \bar Z&=&Y;\\\label{condition}
\bar Z&\in&\mathds{N}_0.
\end{eqnarray}
\end{prop}

To make sure in correctness of the above proposition it is instructive to use the analogy of cyclic sequences with the periodic trajectories on the de Bruijn graph. The $j$'th entry ($j=1,\dots,1+\ell^{p+1}$) of the vector $\bar Z$ contains information of the number of times the trajectory passes through the vertex $e^{(p+1)}_j$. Among all trajectories we choose only the ``smooth'' one's, i.e. those for which the balance conditions hold: the incoming and outgoing ``currents'' in each vertex must be equal. This gives
\begin{equation}\label{closeness}
\bar Z^\dag R^\dag_p=L_p \bar Z.
\end{equation}
The system of linear equations~(\ref{closeness}) has to be supplemented by the initializing conditions, $Y=R_p \bar Z$. It and~(\ref{closeness}) generate the set of equations~(\ref{eq1}) and~(\ref{initialization}). The obtained equations form a system of linear Diophantine equations subject to the condition that solutions $\bar Z$ has to have non-negative entries, i.e. $\bar Z\in \mathds{N}_0$, eq.~(\ref{condition}). 

To resolve the obtained system~(\ref{eq1})-(\ref{condition}) we use the basis of wavelets, constructed by analogy with the basis of \p-adic wavelets~\cite{K2002, KKS2014}. To build the basis vectors we utilize the (non-normalized) basis $\set{\chi_j}_{j=0,\dots,\ell-1}$ of discreet Fourier transform. It is orthogonal on the space $V_\ell$, the vectors' elements are
$$
[\chi_j]_i=e^{\frac{2\pi \i}{\ell}(i-1)j},\qquad i=1,\dots,\ell,\qquad j=0,\dots,\ell-1.
$$
Its orthogonality follows from the nullification of the sum $\sum_{k=0}^{\ell-1} e^{\frac{2\pi \i}{\ell}k}=0$. The vectors of the wavelet basis in the space $V_{\ell^{p}}$ for $p\ge 2$ have the form
\begin{equation}\label{wavelet0}
\psi^{(p)}_{0,j}=\ell^{-p/2}\chi_j \otimes \bigotimes_{k=2}^p q,\qquad j=0,\dots,\ell-1;
\end{equation}
\begin{multline}\label{wavelet1}
\qquad\qquad\qquad\psi^{(p)}_{\gamma,j,\set{\alpha_k}}=\ell^{(p-\gamma)/2}\bigotimes_{k=1}^{\gamma} e_{\alpha_k} \otimes \chi_j \otimes \bigotimes_{k=\gamma+2}^p q,\\ \gamma=1,\dots,p-1\quad j=1,\dots,\ell-1,\quad \alpha_k=0,\dots,\ell-1.
\end{multline}
The total number of vectors is $\ell^p$, their orthogonality follows from orthogonality of vectors $e_k$ and the mutual orthogonality of $\chi_j$ and $q$.

Consider now the action of operators $L_p$ and $R_p$ on the above basis vectors:
\begin{equation}\label{Rp1}
R_p\psi^{(p+1)}_{p,j,\set{\alpha_k}}=0,\qquad
R_p\psi^{(p+1)}_{\gamma,j,\set{\alpha_k}}=\sqrt{\ell}\;\psi^{(p)}_{\gamma,j,\set{\alpha_k}},\qquad \gamma=0,\dots,p-1;
\end{equation}
\begin{equation}
 L_p \psi^{(p+1)}_{0,0}=\sqrt{\ell}\; \psi^{(p)}_{0,0};\qquad  L_p \psi^{(p+1)}_{0,j}=0,\qquad j=1,\dots,\ell-1;
\end{equation}
\begin{equation}
 L_p \psi^{(p+1)}_{\gamma,j,\set{\beta,\alpha_k}}= \psi^{(p)}_{\gamma-1,j,\set{\alpha_k}},\quad \gamma=1,\dots,p-1,\quad \beta=0,\dots,\ell-1,\quad 1\le k\le\gamma-1.
\end{equation}

It follows from~(\ref{initialization}) and~(\ref{Rp1}) that the first $\ell^p$ coefficients (corresponding to $\gamma=0,\dots, p-1$) in the wavelet expansion of $\bar Z$, 
$$
\bar Z=\sum_{j=1}^\ell C_{0,j}\psi^{(p+1)}_{0,j}+\sum_{\gamma=1}^p\sum_{j=1}^{\ell-1}\sum_{\set{\beta, \alpha_k}} C_{\gamma,j,\set{\beta,\alpha_k}}\psi^{(p+1)}_{\gamma,j,\set{\beta,\alpha_k}},\qquad 1\le k\le \gamma-1,
$$
 coincide, up to the factor $\sqrt{\ell}$, with the corresponding coefficients $\tilde C_{\gamma,j,\set{\alpha_k}}$ in the expansion of $Y$,
\begin{equation}\label{Ctilde}
\tilde C_{\gamma,j,\set{\alpha_k}}\equiv\big( \psi^{(p)}_{\gamma,j,\set{\beta,\alpha_k}}\big)^\dag\cdot Y=\sqrt{\ell}C_{\gamma,j,\set{\beta,\alpha_k}},\qquad \gamma=0,\dots,p-1.
\end{equation}
On the other hand since  $Y=L_p \bar Z$ we derive
\begin{multline}\label{LpZbar}
Y=L_p \bar Z=\sqrt{\ell}\;C_{0,0} \psi^{(p)}_{0,0}+\sum_{\gamma=2}^{p-1}\sum_{j=1}^{\ell-1}\sum_{\set{\alpha_k}}\left[\sum_\beta C_{\gamma,j,\set{\beta,\alpha_k}}\right]\psi^{(p)}_{\gamma-1,j,\set{\alpha_k}}\\+\sum_{j=1}^{\ell-1}\sum_{\set{\alpha_k}}\left[\sum_\beta C_{p,j,\set{\beta,\alpha_k}}\right]\psi^{(p)}_{p-1,j,\set{\alpha_k}}.
\end{multline}
The similar relation between the coefficients in wavelet expansions of $X$ and $Y$ results in
\begin{equation}\label{XWave}
X=L_{p-1}Y=\sqrt{\ell}\;\tilde C_{0,0} \psi^{(p-1)}_{0,0}+\sum_{\gamma=2}^{p-1}\sum_{j=1}^{\ell-1} \sum_{\set{\alpha_k}}\left[\sum_\beta \tilde C_{\gamma,j,\set{\beta,\alpha_k}}\right]\psi^{(p-1)}_{\gamma-1,j,\set{\alpha_k}}.
\end{equation}
Notice, that substitution of the coefficients~(\ref{Ctilde}) into the latter equation~(\ref{XWave}) and simultaneous direct multiplication of~(\ref{XWave}) on $q$ results in
$$
X\otimes q=\ell\;\sqrt{\ell}\;C_{0,0} \psi^{(p)}_{0,0}+\ell\;\sum_{\gamma=2}^{p-1}\sum_{j=1}^{\ell-1} \sum_{\set{\alpha_k}}\left[\sum_\beta C_{\gamma,j,\set{\beta,\alpha_k}}\right]\psi^{(p)}_{\gamma-1,j,\set{\alpha_k}}.
$$
It, together with~(\ref{LpZbar}), gives the set of equations for the $(\ell-1)\ell^p$ unknown coefficients $C_{p,j,\set{\beta,\alpha_k}}$
\begin{equation}\label{maineq}
Y-X\otimes q/\ell=\sum_{j=1}^{\ell-1}\sum_{\set{\alpha_k}}\left[\sum_\beta C_{p,j,\set{\beta,\alpha_k}}\right]\psi^{(p)}_{p-1,j,\set{\alpha_k}},
\end{equation}
By analogy wit the above calculations, from~(\ref{Ctilde}) we obtain the set of constraints for the coefficients $C_{p,j,\set{\beta,\alpha_k}}$
\begin{equation}\label{mainconstr}
Y\otimes q/\ell+\sum_{j=1}^{\ell-1}\sum_{\set{\beta,\alpha_k}}C_{p,j,\set{\beta,\alpha_k}}\psi^{(p+1)}_{p,j,\set{\beta,\alpha_k}}\in \mathds{N}_0.
\end{equation}

The system of equations~(\ref{maineq}),~(\ref{mainconstr}) is the result of the first step of the algorithm. The advantage of such reformulation of the original set of Diophantine equations~(\ref{eq1}),~(\ref{initialization}) and~(\ref{condition}) is that now the whole set of equations, due to the special structure of wavelet basis, is divided by $(\ell-1)\ell^{p-1}$  systems of $\ell-1$ linearly independent equations on $\ell$ variables containing also $\ell^2$ conditions. Notice, that in case when $n<\ell^p$ one has to solve maximally $n$ system of equations, while the rest automatically generates trivial solutions only. For composite $n$ solution of equations~(\ref{maineq}),~(\ref{mainconstr}) corresponding to the sequence with non-trivial period $d$ in case when $n/d<\ell^p$ results in vector $\bar Z$, which non-zero components are all equal to $d$ and their total number is $n/d$. 

\begin{ex}In case $\ell=4$ the vector $\psi^{(p)}_{p-1,j,\set{\alpha_k}}$ contains only four non-zero elements. There positions within the vector are marked by the indices $m= 1+\sum_{k=1}^p\alpha_k 4^k$, $m+1$, $m+2$, $m+3$. After simplification of the left-hand-side of~(\ref{maineq}) for each $m$ we arrive to three linearly independent equations for four complex variables $c_{m}$, $c_{m+1}$, $c_{m+2}$, $c_{m+3}$:
\begin{eqnarray*}
Y_{m}-3Y_{m+1}+Y_{m+2}+Y_{m+3}&=&-2\left(c_{m}+c_{m+1}+c_{m+2}+c_{m+3}\right)\\
Y_{m}+Y_{m+1}-3Y_{m+2}+Y_{m+3}&=&2\left(c_{m}+c_{m+1}+c_{m+2}+c_{m+3}\right)\\
Y_{m}+Y_{m+1}+Y_{m+2}-3Y_{m+3}&=&-2\left(c_{m}+c_{m+1}+c_{m+2}+c_{m+3}\right)
\end{eqnarray*}
Among all solutions of the above system one has to choose those that produce non-negative integer values of $\bar Z$ calculated from the formulae ($j=0,1,2,3$):
\begin{eqnarray*}
\bar Z_{m+j 4^p}&=&\frac{Y_{m+j 4^p}}{4}+\frac{3 c_{m+j}}{2};\\
\bar Z_{m+j 4^p+1}&=&\frac{Y_{m+j 4^p}}{4}-\frac{ c_{m+j}}{2};\\
\bar Z_{m+j 4^p+2}&=&\frac{Y_{m+j 4^p}}{4} + \frac{c_{m+j}}{2};\\
\bar Z_{m+j 4^p+3}&=&\frac{Y_{m+j 4^p}}{4}-\frac{ c_{m+j}}{2}.
\end{eqnarray*}
\end{ex}
The case $\ell=2$ is considered in details in section 3.4.

\subsection{The second step of the algorithm.}
The system of $\ell^p$ linear equations~(\ref{maineq}),~(\ref{mainconstr}) contains $\ell^{p+1}$ variables and has more then one solution. Some of them can correspond to disjoint trajectories on the graph. Therefore the second step of the algorithm:  ``taking the intersection operation in $\set{\bar Z}\cap \mathcal{F}_{p+1,n}$'' means separation of the disjoint trajectories from the closed ones. Let us, first, reproduce the analytic non-linear criterion suggested in~\cite{GO2013a}. 

\begin{prop} Let  $\bar Z\cdot\Phi$ denote scalar multiplication of vector $\bar Z$ and vector $\Phi=[\phi_1,\dots, \phi_{\ell^{p+1}}]$, let $\Lambda_{\Phi}$ is a diagonal matrix of phases $\mathrm{diag}\set{e^{\i \phi_1}\dots e^{\i\phi_{\ell^{p+1}}}}$, and $A_{\ell^{p+1}}[\bar Z]$ is a matrix obtained by multiplication $L_{p+1}\cdot\mathrm{diag}\set{\bar Z}\cdot R_{p+1}$, then $\bar Z$ corresponds to a cyclic sequence iff
the integral
\begin{equation}\label{integral}
I\equiv\frac{1}{n}\prod_{i=1}^{\ell^{p+1}}\left[\int_{[0,2\pi]}\frac{d\phi_i}{2\pi}\right]\cdot e^{-\i\;\bar Z\cdot\Phi}\;\tr \big(A_{\ell^{p+1}}[\bar Z]\Lambda_{\Phi}\big)^{n},
\end{equation}
is positive.
\end{prop}

Indeed, the trace of matrix raised to the power $n$ is a sum of all possible products of $n$ matrix elements taken along the legitimate closed paths on the sub-graph isolated from the frame by the special choice of the adjacency matrix $A_{\ell^{p+1}}[\bar Z]$. The phases marks the individual paths on the sub-graph and the integral nullifies those that passes not all edges. Moreover, the non-zero part of the integral, $I$, gives the number of $(p+1)$-close cyclic sequences, which correspond to the frequency vector $\bar Z$. this proofs the proposition 2.

The multidimensional integral~(\ref{integral}) is a good object for analytic study~\cite{GO2013a, GO2013b}, while the practical calculation would require significant amount of resources. In applications it can be replaced by the direct test of the matrix $A_{\ell^{p+1}}[\bar Z]$. The trajectory is closed if the corresponding sub-graph is connected, i.e. the adjacency matrix $A_{\ell^{p+1}}[\bar Z]$ consists of only one non-zero block. Matrix $A_{\ell^{p+1}}[\bar Z]$ has a special properties, which simplifies the testing:
\begin{itemize}
\item If $\bar Z$ contains no zeros, then the trajectory passes through all vertices of the graph $G_\ell(p)$ and the graph is obviously connected.
\item Each row and each column of the matrix  $A_{\ell^{p+1}}[\bar Z]$ contains not more then $\ell$ non-zero elements or does not contain them at all. To demonstrate it let us represent the vector $\bar Z$ as a direct sum of its $\ell$ parts, such that first $\ell$ entries belong to the first part, the second $\ell$ entries belongs to the second part and so on ..., i.e. $\bar Z=\oplus_{j=1}^\ell\bar Z_j$. The total size of non-zero blocks of the matrix  $A_{\ell^{p+1}}[\bar Z]$ coincides with the number of non-zero entries of the vector $\sum_{j=1}^\ell\bar Z_j$. Other vertices of the graph $G_\ell(p)$ are not visited. Thus the total size of the non-zero matrix block (blocks) is not larger then~$n$.
\end{itemize}

Presence of a single block in the matrix $A_{\ell^{p+1}}[\bar Z]$ on the graph's language means existence of at least one spanning tree~\cite{KR2000}. Thus one can either check the non-nullification of the cofactor, see eq.~(\ref{cofactor}), constructed for the graph with the adjacency matrix $A_{\ell^{p+1}}[\bar Z]$, or rearrange the vertices (permute the rows and columns in the matrix) to reveal the block structure explicitly. The latter strategy, out of its effectiveness in case of sparse matrices, has been utilized in the numerical construction of the clusters presented on fig.~\ref{Fig5}.

\subsection{The case $\ell=2$.}The known generic results regarding the cluster structure are gathered in section~2.4. Here we discuss a few more observations for the case $\ell=2$.

\begin{figure}[t]
\begin{center}{
\includegraphics[height=5.5cm]{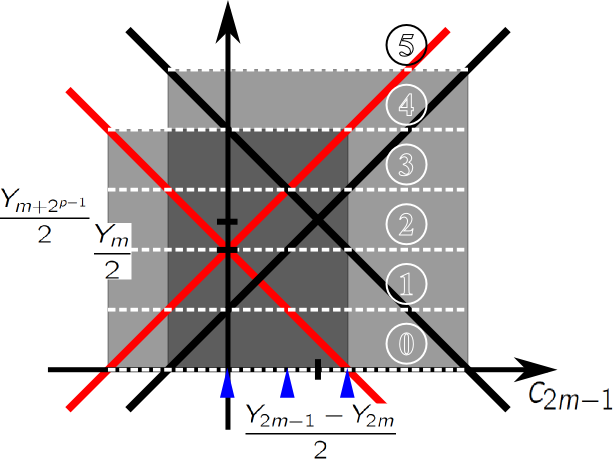}
}\end{center}
\caption{ \small Scheme for solution of equations~(\ref{Y1})-(\ref{Y2}). The functions $\frac{Y_{m}}{2}\pm c_{2m-1}$ are plotted in red, the black lines are functions $\frac{Y_{m+2^{p-1}}}{2}\pm\left(\frac{Y_{2m-1}-Y_{2m}}{2}- c_{2m-1}\right)$. The grey rectangles shows the area of non-negative $\bar Z$,  the white dashed lines denote the possible integer values of $\bar Z$. The feasible values of $c_{2m-1}$ (see blue marks) can be found as coordinates of intersections of the integer-valued lines with the red and black lines within the intersection of grey areas (dark grey area).
}\label{Fig2} 
\end{figure}

The set of equations~(\ref{maineq}),~(\ref{mainconstr}) in case $\ell=2$ takes the form
\begin{eqnarray}\label{Y1}
\frac{Y_{2m-1}-Y_{2m}}{2}&=&c_{2m-1}+c_{2m};\\
\bar Z_{2m-1}=\frac{Y_{m}}{2}+c_{2m-1}&\in &\mathds{N}_0;\\
\bar Z_{2m}=\frac{Y_{m}}{2}-c_{2m-1}&\in &\mathds{N}_0;\\
\bar Z_{2m+2^p-1}=\frac{Y_{m+2^{p-1}}}{2}+c_{2m}&\in &\mathds{N}_0;\\
\bar Z_{2m+2^p}=\frac{Y_{m+2^{p-1}}}{2}-c_{2m}&\in &\mathds{N}_0.\label{Y2}
\end{eqnarray}
For convenience we introduced the running index $m=1,\dots,2^{p-1}$. Let us consider level $p=1$, here the frequency vector can be represented as  $Y^{(z,p=1)}=[n-z,z]$. The symmetry of clusters allows us to define $z$ to be less then the integer part of $\frac{n}{2}$, we also drop the trivial clusters, $z=0,1$. Parameter $z$ determines the number branches at the level $p=1$ (see for example tree on the fig.~\ref{Fig5}). Indeed, one can check that the number of solutions of equations~(\ref{Y1})-(\ref{Y2}) corresponding to the closed graphs is equal to $z$, see fig~\ref{Fig2} and the text in the caption. For instance,
$$
\begin{array}{l|cccc}
z\diagdown c_1& \frac{n-z}{2}-1& \frac{n-z}{2}-2& \frac{n-z}{2}-3& \frac{n-z}{2}-4 \\\hline
2&c_2=0&c_2=1\\
3&c_2=-\frac{1}{2}&c_2=\frac{1}{2}&c_2=\frac{3}{2}\\
4&c_2=-1&c_2=0&c_2=1&c_2=2\\
\vdots
\end{array}
$$

Our numerical study of binary cyclic sequences (see for example fig.~\ref{Fig5}) shows that for prime $n$ the maximal level, where the tree is branching, is reached in the cluster starting from the frequency vector $Y^{(3,1)}=[n-3,3]$. This universal property can be explained within the proposed method of wavelet analysis by the following arguments. On the $p$th step of hierarchy one has to solve the system~(\ref{Y1})-(\ref{Y2}). On average all non-zero elements of $\bar Z^{(z,p)}$ are half of the elements of $Y^{(z,p)}$. This principle is violated if the difference between the nearest elements, $\frac{Y_{2m-1}^{(z,p)}-Y_{2m}^{(z,p)}}{2}$, is large. An additional requirements is that the sequence should repeat itself as many times as possible, i.e. to have encounter of the largest length and order. Therefore the second scenario, when $c_1= \frac{n-z}{2}-2$, is the preferable one. On the level $p=2$ the resulting frequency vectors are $Y^{(z,2)}=[n-z-2,2,2,z-2]$. The procedure can be iterated till the level $p=\floor{\frac{n-z}{2}}$. Thus $z=2$ and $z=3$ give the maximal number of iterations. The obtained after the iterations vectors are $[3,2,2,0,\dots]$ for $z=2$  and $[2,2,1,1\dots]$ for $z=3$. The latter vector still contains one more encounter, it corresponds to two choices of symbolic sequences: $[\dots 0\,\,1\,0\,1\,0\dots]$ and $[\dots 0\,\,1\,1\,0\,0\dots]$ (note, that geometrically the first entry of $Y$ means the number of visits of the loop $00\dots 0$ on the graph $G_2(p)$, i.e. the left loops on the graphs plotted on fig.~\ref{Fig6}). Thus the maximal level of hierarchy $p_{max}$ is achieved in the cluster marked by $Y^{(3,1)}=[n-3,3]$ and it is equal to $p_{max}=\floor{\frac{n-3}{2}}+1$, $p_{max}=5$ for $n=11$, see fig.~\ref{Fig5}.

\subsection{Two-fold de Bruijn sequences.}
Consider the problem of finding the number, $\mathcal{N}$ of two-fold de Bruijn sequences (see definition 2) for the binary alphabet.  In this section we propose a generic formula for counting of two-fold de Bruijn sequences.
\begin{prop}Let function $\Phi_{2^p}(2k)$ counts the number of connected graphs obtained from de Bruijn graph $G_2(p)$ by removing $k$ pairs of edges, such that the removed edges are marked by pair of the basis vectors $\set{e_{2m-1}^{(p+1)},e_{2m+2^p-1}^{(p+1)}}$ or $\set{e_{2m}^{(p+1)},e_{2m+2^p}^{(p+1)}}$, where $m$ takes $k$ different values from the set of integers $1,\dots,2^{p-1}$. The number, $\mathcal{N}_p$ of $p$-ary two-fold de Bruijn sequences can be calculated by the formula  
\begin{equation}\label{MainFormula}
\mathcal{N}_p=2^{2^{p-1}-p}+\sum_{k=1}^{2^{p-1}-1}\frac{2^{2k-\beta}}{4k} \prod_{j=0}^{\beta/2-1}\left(1+4\cos^2(\frac{2\pi}{\beta} j)\right)\cdot\Phi_{2^p}(2^p-2k)+2^{2^{p}-p-1},
\end{equation}
where $\beta= 2k-1-\frac{k}{2^{\gamma_0}}\sum_{j=0}^{\gamma_0}2^j$ and $\gamma_0$ is the multiplicity of the prime factor 2 in the integer index $k$.
\end{prop}

Below we give approval of the formula~(\ref{MainFormula}) and analyze its ingredients in more details. 

Any $p$-ary two-fold de Bruijn sequence can be seen as a closed trajectory on graph $G_2(p-1)$, where each of $2^p$ edges is passed exactly two times. All entries of the corresponding frequency vector $Y\in\mathcal{F}_{p,2^{p+1}}$ are equal to $2$. Then for each $m$ ($m=1,\dots, 2^{p-1}$) one gets from~(\ref{Y1})-(\ref{Y2}) the system of equation
\begin{eqnarray}\label{c1}
c_{2m-1}+c_{2m}&=&0;\\
1+c_{2m-1}&\in &\mathds{N}_0;\\
1-c_{2m-1}&\in &\mathds{N}_0;\\
1+c_{2m}&\in &\mathds{N}_0;\\
1-c_{2m}&\in &\mathds{N}_0.\label{c2}
\end{eqnarray}
The systems~(\ref{c1})-(\ref{c2}) generate  only three admissible solutions, they are $c_{2m-1}=c_{2m}=0$, $c_{2m-1}=-c_{2m}=1$, $c_{2m-1}=-c_{2m}=-1$.  Thus the corresponding values of the vector $\bar Z$ are
\begin{eqnarray}
\label{sol1}&&\bar Z_{2m-1}=1,\quad\bar Z_{2m}=1,\quad\bar Z_{2m+2^p-1}=1,\quad\bar Z_{2m+2^p}=1\\
\label{sol2}&&Z_{2m-1}=2,\quad\bar Z_{2m-1}=0,\quad\bar Z_{2m+2^p-1}=0,\quad\bar Z_{2m+2^p}=2\\
\label{sol3}&&\bar Z_{2m}=0,\quad\bar Z_{2m}=2,\quad\bar Z_{2m+2^p-1}=2,\quad\bar Z_{2m+2^p}=0.
\end{eqnarray} 
The total number of elements in the set $\set{\bar Z}$ is $\norm{\set{\bar Z}}=3^{2^{p-1}}$. Having obtained the solutions, first, one has to select from them the legitimate ones, i.e. those that produce connected graphs with the connectivity matrix $A_{2^{p+1}}[\bar Z]$ (see section 3.3 for definition). Second, one has to calculate the number of Eulerian cycles on the obtained graphs. Note that some edges are doubled ($\bar Z$ contains $2$ at corresponding position) and the counting of the Eulerian cycles with the help of BEST theorem, eq.~(\ref{EulerCycles}), aliquot increases the number of cycles, while the corresponding sequences are indistinguishable. To avoid this miscounting one can instead calculate the Eulerian cycles on the graph minors obtained by contracting the doubled edges. The structure of the matrix $A_{2^{p+1}}[\bar Z]$ (see analysis in section 3.3 and also example in section 3.5) allows to conclude that the adjacency matrices of such minors depends only on the total number, $k$, of solutions~(\ref{sol2}) and~(\ref{sol3}) entering the vector $\bar Z$ and can be represented in a generic form. The $2k\times 2k$ adjacency matrix $Q'_{2k}$ of the graph $G_2(p)$ minor formed by contracting of $2^p-2k$ edges is defined by the formula
\begin{equation}\label{Q2k}
Q'_{2k}=q^\dag\otimes\1_k\otimes q,
\end{equation}
where $q$ is the column vector of ones, $q^\dag=(1,1)$ (compare with~(\ref{Qellp})). Thus one can divide the set  $\set{\bar Z}$ by $2^{p-1}$ subsets with respect to the number of  solutions~(\ref{sol2}) and~(\ref{sol3}) entering the vectors. Therefore the number of two-fold de Bruijn sequences is a sum over $k=0,\dots,2^{p-1}$ of the cofactors, defining the number of Eulerian cycles, multiplied by the function $\Phi_{2^p}(2^p-2k)$  counting the number of connected graphs,
$$
\mathcal{N}_p=\sum_{k=0}^{2^{p-1}}Co_1[2\1_{2k}-Q'_{2k}]\cdot\Phi_{2^p}(2^p-2k). 
$$  
 
It is natural to single out the terms corresponding to $k=0$ and $k=2^{p-1}$. The cofactor in case $k=0$ should be explicitly defined as equal to one. Thus the only contribution is given by the function $\Phi_{2^p}(2^{p-1})$. When all edges are doubled each of the corresponding two-fold de Bruijn sequences is simply a combination of two copies of the single-folded de Bruijn sequences, i.e.  $\Phi_{2^p}(2^p)=2^{2^{p-1}-p}$ is the number of Eulerian cycles on the graph $G_2(p-1)$ or the number of Hamiltonian cycles on $G_2(p)$. The last term,  $k=2^{p-1}$, can be estimated in the similar way. Indeed, here the trajectory fully covers the graph $G_2(p)$, thus  $\Phi_{2^p}(0)=1$ and the cofactor gives the number of Eulerian cycles on  $G_2(p)$, which is $2^{2^{p}-p-1}$ and we arrive to
\begin{equation}\label{MainFormula2}
\mathcal{N}_p=2^{2^{p-1}-p}+\sum_{k=1}^{2^{p-1}-1}Co_1[2\1_{2k}-Q'_{2k}]\cdot\Phi_{2^p}(2^p-2k)+2^{2^{p}-p-1}. 
\end{equation}
  
Finally, we derive explicit formula for the cofactor in~(\ref{MainFormula2}). The secular equation for the matrix $Q'_{2k}$ given by~(\ref{Q2k}) has the form
\begin{equation}\label{Cof}
\det[\lambda -Q'_{2k}]=(\lambda-2)\lambda^{2k-1} \prod_{j=0}^{\beta-1}\left(1-\lambda^{-1} e^{\frac{2\pi \i}{\beta} j}\right),
\end{equation}
where $\beta= 2k-1-\frac{k}{2^{\gamma_0}}\sum_{j=0}^{\gamma_0}2^j$ and parameter $\gamma_0$ is the multiplicity of the prime factor 2 in the natural number $k$.
Indeed, the eigenvalue $2$ of the matrix $Q'_{2k}$ corresponds to the constant vector, the $2k-\beta-1$ zero eigenvalues is the reflection of the projection properties of the matrix. It projects the span of the $2k-\beta-1$ vectors  from the wavelet basis~(\ref{wavelet1}) with $\gamma\ge\gamma_0$ onto the complementary subspace. In this subspace the matrix $Q'_{2k}$ becomes a matrix of cyclic permutations. 

To obtain expression for the sought cofactor one can use the following formula~\cite{R2002}
\begin{equation}\label{cof}
Co_1[2\1_{2k}-Q'_{2k}]=\frac{1}{2k}\left.\frac{d}{d\lambda}\det[\lambda -Q'_{2k}]\right|_{\lambda=2}=\frac{2^{2k-\beta}}{4k} \prod_{j=0}^{\beta/2-1}\left(1+4\cos^2(\frac{2\pi}{\beta} j)\right)
\end{equation}
This finishes the proof of the proposition 3.

The function $\Phi(2k)$ can be represented as a sum over permutations of $k$ solutions~(\ref{sol2}) and~(\ref{sol3}) in the vector $\bar Z$. Thus
$$
\Phi_{2^p}(2^p-2k)=\sum_{\alpha=0}^{2^{p-1}-k} \sum_{\sigma(\alpha,k,2^{p-1})} \Theta\left(A_{2^{p+1}}[\bar Z_\sigma]\right),
\quad
\Theta\left(A\right)=\begin{cases}1,&\mbox{ $A$ has only one block;}\\
0,&\mbox{ $A$ has more then one block.}
\end{cases}
$$
where $\sigma(\alpha,k,2^{p-1})$ denotes the permutations of three types of solutions~(\ref{sol1})-(\ref{sol3}) entering $\bar Z$ over $2^{p-1}$ places, among them there are $\alpha$ solutions of the sort~(\ref{sol2}) and $2^{p-1}-k-\alpha$ of the sort~(\ref{sol3}). $\bar Z_\sigma$ denotes the vector from $\set{\bar Z}$ with particular permutation. 

The number of permutations of $\alpha$ and $2^{p-1}-k-\alpha$ elements over $2^{p-1}$ positions is known to be 
$$
\mathrm{PermNo}=\sum_{\alpha=0}^{2^{p-1}-k} \sum_{\sigma(\alpha,k,2^{p-1})} 1=\sum_{\alpha=0}^{2^{p-1}-k} \frac{(2^{p-1})!}{\alpha!k!(2^{p-1}-k-\alpha)!}=2^{2^{p-1}-k}\frac{(2^{p-1})!}{k!(2^{p-1}-k)!}.
$$ 
This is the number of graphs to be tested. It can be reduced by removing solutions which has $2$ at the first or/and at the last positions of the vector $\bar Z$, since they correspond to disconnected graphs. Analysis based on the similar combinatiorial arguments and the specifics of de Bruijn graphs allows to find a few more explicit expressions for the function  $\Phi_{2^p}(2^p-2k)$, they are: $\Phi_{2^p}(2^p-2)=2^{2^{p-1}-1}$, $\Phi_{2^p}(2)=2^p-2$, $\Phi_{2^p}(4)=2(2^{p-1}-1)(2^{p-1}-2)-\delta_{p,3}$. 

Below we calculate explicit numbers of two-fold de Bruijn sequences by the formula~(\ref{MainFormula}) for $p=2,3,4$. For $p=1$ there are only two two-fold de Bruijn sequences: $[1,1,0,0]$ and $[1,0,1,0]$. Consider the case $p=2$. In this case we have to analyze only 4 vectors from the set $\set{\bar Z}$, they are
\begin{multline*}
\set{\bar Z}=\left\{[1,1,1,1,1,1,1,1],\right.\\\left. [1,1,0,2,1,1,2,0],\quad [0,2,0,2,2,0,2,0],\quad
[0,2,1,1,2,0,1,1]\right\}.
\end{multline*}
Note, that other vectors, which contain $2$ at the first and/or the last positions are thrown out (see the previous paragraph). The first vector $\bar Z=[1,1,1,1,1,1,1,1]$ generates path on the full graph $G_2(3)$. There are two Eulerian paths on this graph, which correspond to the sequences $[1,1,1,0,0,0,1,0]$ and $[1,1,1,0,1,0,0,0]$. The third vector $\bar Z=[0,2,0,2,2,0,2,0]$ is a composition of two identical two-ary single-fold de Bruijn sequences, i.e. the sequence $[1,1,0,0,1,1,0,0]$. Two other vectors correspond to the sub-graphs with the following adjacency matrices
\begin{equation*}
A_{4}[1,1,0,2,1,1,2,0]=\left(\begin{array}{cccc}
1&0&1&0\\
1&0&1&0\\
0&0&0&2\\
0&2&0&0
\end{array}\right)\quad A_{4}[0,2,1,1,2,0,1,1]=\left(\begin{array}{cccc}
0&0&2&0\\
2&0&0&0\\
0&1&0&1\\
0&1&0&1
\end{array}\right).
\end{equation*} 
By contracting the doubled edges we obtain the minors of $G_2(3)$. In both cases the sub-graph's minors are described by one and the same adjacency matrix
$$
\left(\begin{array}{cc}
1&1\\
1&1
\end{array}\right)
$$
The number of Eulerian cycles on the minors can be calculated by the formula~(\ref{cof}), it equals one. Thus we obtain $\mathcal{N}_2=5$. The two more sequences are $[1,1,0,0,0,1,1,0]$, $[1,1,1,0,0,1,0,0]$. 

The results for $p=3,4$ are gathered in the tables below
\begin{center}
   \begin{tabular}{ l }\\
   \begin{tabular}{ l  || c  | c  | c  | c | c }
   \multicolumn{6}{l}{$p=3$, $\mathcal{N}_3=72$} \\ \hline
     $k$                & 0  & 1  & 2  & 3 & 4 \\ \hline\hline
     $\mathrm{PermNo}$  & 16 & 32 & 24 & 8 & 1\\ \hline
     $\Phi_{8}(8-2k)$ & 2  & 8  & 11 & 6 & 1\\
    \hline
$Co_1[2\1_{2k}-Q_{2k}]$ & 1  & 1  & 2  & 4 & 16
  \end{tabular} \\\\
   \begin{tabular}{ l  || c  | c  | c  | c | c | c | c | c | c }
   \multicolumn{6}{l}{$p=4$, $\mathcal{N}_4=43\,768$} \\ \hline
     $k$                & 0  & 1  & 2  & 3 & 4 & 5 & 6 & 7 & 8 \\ \hline\hline
     $\mathrm{PermNo}$  & 256 & 1024 & 1792 & 1792 & 1120 & 448 &112 & 16 & 1\\ \hline
     $\Phi_{16}(16-2k)$ & 16 & 128 & 380 & 584 & 519 & 274 & 84 & 14 & 1\\
    \hline
$Co_1[2\1_{2k}-Q_{2k}]$ & 1  & 1  & 2  & 4 & 16 & 48 & 128 & 448 & 2048
  \end{tabular}
  \end{tabular}
\end{center}

\section{Discussion}\label{Sec4}
In present article we propose a method for studying of cluster structure of the set of cyclic symbolic sequences. The main analytic tool we use is analysis of properties of the operators acting as raising and lowering operators on the cluster tree. Formalization of the operators allowed to find a proper basis, the wavelet basis, for optimization of the cluster analysis procedure and reduce the problem to two-stage algorithm including solution of linear Diophantine equations and analysis of block structure of matrices. 

The proposed approach allowed to extend a number of solvable problems in combinatorial mathematics. In particular, we presented a formula~(\ref{MainFormula}) for counting number of \p-ary two-fold de Bruijn sequences. Note here, that the formula~(\ref{MainFormula}) includes the counting function $\Phi_{2^p}(2^p-2k)$, which explicit form was found only for a few values of $k$, some other values were tabulated in the previous section. We believe, however, that its behavior and asymptotic can be further studied with the help of the method of generating functions, widely used in combinatorics~\cite{St1999}. 

The known generic results regarding the cluster structure of symbolic sequences are gathered in section~2.4. Our study allowed us to add the estimation of the maximal branching level of the tree, it is $p_{max}=\floor{\frac{n-3}{2}}+1$. Though this result was rigorously obtained for prime $n$ only, there is no obstacles to get the same result for all $n$ by the same reasoning, our numeric study shows universality of this result.   

The wavelet analysis of cyclic symbolic sequences can be easily extended onto the case of open sequences of a finite length. By introducing an additional letter and closing the sequence by a long arc made up of these letters we can map the problem of open sequences back onto the problem of cyclic sequences. Certainly the algorithms should be improved to take into account the artificial part of the sequence.

We believe that the approach based on the wavelet analysis opens new possibilities for optimization of the DNA sequence assembly algorithms. First, the wavelet transform of the frequency vectors allows to reformulate the problem in a way suitable for parallelization of the restoring procedures. Moreover, in the frequency representation of symbolic sequences one has to save only addresses of the non-zero elements of the vector without saving the information about connectivity of the edges on de Bruijn graph, which allows to optimize the data storage problem. Note that the similar ideas, of representing de Bruijn graph as a tree, was also discussed in literature~\cite{CLJSM2014, N2014}.    

In the article we considered only the processes which conserve the length of the sequences, as soon as it is dictated by applications in theory of quantum chaos. In fact, the lowering operator, i.e. solution of Diophantine equations, can be considered without the second part where one controls the connectivity of the sequence. Then the resulting frequency vectors at $p>n/2$ should  corresponds to a set of cyclic sequences, which total length is conserved, rather then to a single sequence. On the contrary, the raising operator acting on such frequency vectors can be seen as a gluing operator acting on several sequences. In such a way one can organize a stochastic process on a set of symbolic sequences. The process can be described as follows, after several subsequent rising of a starting long sequence and the same number of lowerings of the obtained frequency vector with a random choice of possible solutions one can get a solution which instead of one cyclic sequence contains many of them of a shorter lengths. This procedure of rising and lowering repeated several times can be considered as a stochastic process of "random bubbling of symbolic sequences".  

To summarize, we can say that, generically, the idea of classification of sequences with respect to the local content is rather natural. First step in this direction was done in the work of Hamming~\cite{H1950}, who introduced a {\it metric} distance between sequences. The Hamming distance between two given sequences, such that both have the same length and a fixed starting point, is defined as the total number of different symbols appearing in the  identical positions. The metric space generated by the Hamming distance on the set of binary sequences is known as Hamming cube. The Hamming distance, however does not take into account the surrounding symbols. On the contrary, the ultrametric distance is absolutely insensitive to the absolute position of symbols within the sequence, while it takes into account the local surrounding. Thus even one different symbol in two sequences put them on the furthest distance from each other. This prompts us the next natural step in the direction of classification of symbolic sequences. This should be combination of both approaches together. Note that first attempts of such kind are based not on the construction of distances, but on introduction of stochastic processes which take into account both positions of symbols and their local surrounding see the work~\cite{GO2011} and also works~\cite{AIMN2014, MAL2005}. \newline

{\bf Aknowlegements:}
Author thanks S.Nechaev, P.Braun, B.Gutkin, T.Guhr for useful discussion at the beginning of this work. the work was partially supported by KAW foundation.

\appendix
\section{Representation of the set $\mathcal{X}_n^{ \mathcal{A}_2}$ on a regularly branching tree.}\label{AppTree}

\begin{figure}[t]
\centering\includegraphics[width=\textwidth]{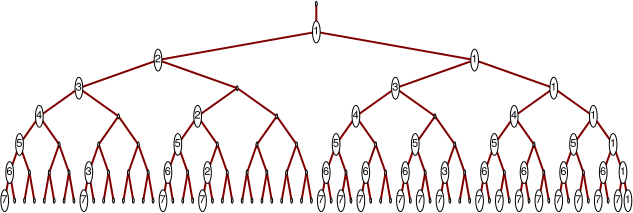}
\caption{\small Graphical representation of the set $X_7^{ \mathcal{A}_2}$ by a binary tree. Only half of the tree (the sequences starting from $1$) is plotted. The representatives of the factor sets $\mathcal{X}_n^{ \mathcal{A}_2}$ ($n=7,6,5,4,3,2,1$) are marked by the circles on corresponding level of the tree.  The nubers inside denote period of the corresponding sequence.}\label{Fig10}
\end{figure}

On the fig.~\ref{Fig10} we graphically represent structure of the factor set $\mathcal{X}_7^{ \mathcal{A}_2}$ by a special choice of its representatives from the  set $X_7^{ \mathcal{A}_2}$. One can compare it with the structure of the same set (see fig.~\ref{Fig5}) revealed by the ultrametric distance~(\ref{ultrdist}). 

The representatives of $\mathcal{X}_7^{ \mathcal{A}_2}$ are chosen to maximize the number $1+\sum_{k=1}^n a_k 2^{n-k}$, i.e. the index of the basis vector~(\ref{trivUltrametricity}). For the case of prime $n$ one can offer a synthetic algorithm based on the following principles:
\begin{itemize}
\item[1.] Sequences are classified by $r=0,\dots,n$, the lengths of the longest sub-string, $L=[1,\dots,1]$, and by the frequency $f$ of appearance $L$ in the sequence. The integer partition of $n$, then has the form
$$
\sum_{j=1}^{s}\nu_j (r+m_j)=n,\qquad \sum_{j=1}^{s}\nu_j=f.
$$
Let $L_j$ be the sequence $L$ followed by arbitrary string of the length $m_j$, as on the scheme 
$$
\underbrace{ \underbrace{1\dots 1}_{r} \underbrace{0\dots 0}_{m_1}}_{L_1} \underbrace{\underbrace{1\dots 1}_{r} \underbrace{0\dots 0}_{m_2}}_{L_2}\dots
$$

\item[2.] Since $n$ is prime, there are at least two non-equal $\nu_j$ entering the integer partition of $n$. Let $\nu_{j_0}=\min_j\set{\nu_j}$, then there is at least one $\nu_{j_1}>\nu_{j_0}$, such that $\nu_{j_1}$ is not divisible by $\nu_{j_0}$. Therefore one can organize a set of cyclic sequences constructed from two letters $L_{j_0}$ and $L_{j_1}$, such that they have no other periods except $\nu_{j_0}+\nu_{j_1}$, for instance
$$
\underbrace{L_{j_0}\dots L_{j_0}}_{\nu_{j_0}}\underbrace{L_{j_1}\dots L_{j_1}}_{\nu_{j_1}}.
$$

\item[3.]  To obtain the representatives one should choose any suitable strategy of ordering of $\nu_{j_1}$ pieces between $\nu_{j_0}$ places, such that the resulting sequences of $L_{j_0}$'s and $L_{j_1}$'s cannot be obtained one from the other by any rotation and, finally, one has to and fill up the yet empty places within the strings $L_j$ by all possible ways.
\end{itemize}

\begin{thebibliography}{99}

\bibitem{K1998} B.P.Kitchens, ``Symbolic Dynamics. 
One-sided, Two-sided and Countable State Markov Shifts'' (Springer Berlin Heidelberg, 1998)

\bibitem{HZ1998} B.-L.Hao, W.-M.Zheng, ``Applied Symbolic Dynamics and Chaos''  (World Scientific, 1998)

\bibitem{DFT2003} C.S.Daw1, C.E.A.Finney, E.R.Tracy ``A review of symbolic analysis of experimental data''
{\it Rev. Sci. Instrum.} {\bf 74} (2003) 915

\bibitem{B1946} N.G.de Bruijn, ``A combinatorial problem'' {\it Indagationes Math.} {\bf 8} (1946) 461

\bibitem{H2009} D.Haussler, S.J.O’Brien, {\it et al.}  ``Genome 10K: a proposal to obtain whole-genome sequence for
10,000 vertebrate species'' {\it J.Hered.} {\bf 100} (2008) 659

\bibitem{GR1961} E. N. Gilbert, J. Riordan, ``Symmetry Types of Periodic Sequences'' {\it Illinois J. Math}{\bf 5} (1961) 657

\bibitem{BH2007} H.Bailin, X.Huimin, ``Factorizable Language Revisited from Dynamics to Biology'' {\it Int. J.Mod.Phys. B} {\bf 21} (2007) 4077

\bibitem{B2000} J.G.Brida, ``Symbolic Time Series Analysis and Economic Regimes'' {\it Structural Change and Economic Dynamics} {\bf 14} (2000) 159

\bibitem{BVKB2010} A.Bolshoy, Z.Volkovich, V.Kirzhner, Z.Barzily, ``Genome Clustering: From Linguistic Models to Classification of Genetic Texts''  (Springer, Heidelberg, 2010) 

\bibitem{T1989} M.Tabor, ``The Surface of Section´´ in ``Chaos and Integrability in Nonlinear Dynamics: An Introduction'' (New York: Wiley, 1989)

\bibitem{O1993} ``Chaos in Dynamical Systems'' (Cambridge University Press, Cambridge, 1993) p.81

\bibitem{CAMTV2012} P.Cvitanovi\'c, R.Artuso, R.Mainieri, G.Tanner, G.Vattay, ``Chaos: Classical and Quantum'' (Niels Bohr Institute, Copenhagen, 2012)

\bibitem{A1987} D.Auerbach, P.Cvitanovi\'c, J.-P.Eckmann, G.Gunaratne, ``Exploring Chaotic Motion Through Periodic Orbits'' {\it Phys.Rev.Letts.} {\bf 23} (1987) 2387

\bibitem{A1990} R.Artuso, E.Aurell, P.Cvitanovi\'c ``Recycling of Strange Sets: I. Cycle Expansions'', {\it Nonlinearity} {\bf 3} (1990) 325;
``Recycling of Strange Sets: II. Applications'', {\it Nonlinearity} {\bf 3} (1990) 361

\bibitem{G1990} M.G. Gutzwiller,  ``Chaos in classical and quantum mechanics'' (Springer, New York, 1990)

\bibitem{SR2001} M.Sieber, K.Richter, ``Correlations between Periodic Orbits and their Role in Spectral Statistics'' {\it  Phys.Scripta.} {\bf T90} (2001) 128  %4

\bibitem{HMBH2004} S.Heusler, S.M\"uller, P.Braun, F.Haake ``Universal spectral form factor for chaotic dynamics'' {\it  J.Phys.A:Math.Gen.} {\bf 37} (2004) L31 %6

\bibitem{H2010} F.Haake,  ``Quantuim signatures of chaos'', 3-d edition (Springer, Heidelberg, 2010) 

\bibitem{GO2013a} B. Gutkin, V.Al. Osipov ``Clustering of periodic orbits in chaotic systems''  {\it Nonlinearity} \textbf{26} (2013) 177

\bibitem{M2008} F.Murtagh ``Identifying and Exploiting Ultrametricity''
in ``Data Analysis, Machine Learning and Applications'' (Springer, 2008) p.263%-272

\bibitem{F1894} C. Flye Sainte-Marie, ``Solution to question nr. 48'' {\it L'interm\'ediaire des Math\'ematiciens} {\bf 1} (1894) 107

\bibitem{CPT2011} P.E.C.Compeau, P.A.Pevzner, G.Tesler ``How to apply de Bruijn graphs to genome assembly'' {\it Nature Biotechnology} {\bf 29} (2011) 987

\bibitem{B1946a} N.G.de Bruijn, ``A Combinatorial Problem'', {\it Koninklijke Nederlandse Akademie v. Wetenschappen} {\bf  49} (1946) 758

\bibitem{G1946} I.J.Good, ``Normal recurring decimals'' {\it Journal of the London Mathematical Society} {\bf 21} (1946) 167

\bibitem{CBP2009} M.J.Chaisson, D.Brinza, P.A.Pevzner ``De novo fragment assembly with short mate-paired reads: Does the read length matter?'' {\it Genome Res.} {\bf 19} (2009) 336

\bibitem{CB2011} T.C.Conway, A.J.Bromage ``Succinct data structures for assembling large genomes'' {\it Bioinformatics} {\bf 27} (2011) 479

\bibitem{PHCHTB2012} J.Pell, A.Hintze, R.Canino-Koning, A.Howe, J.M.Tiedje, C.T.Brown,  ``Scaling metagenome sequence assembly with probabilistic de Bruijn graphs''  {\it Proc. Nat.Ac.Sci.} {\bf 109} (2012) 13272

\bibitem{CLJSM2014} R.Chikhi, A.Limasset, S.Jackman, J.T.Simpson, P.Medvedev, 
``On the Representation of de Bruijn Graphs'' {\it  Lecture Notes in Computer Science} {\bf 8394} (2014) 35

\bibitem{AZ2007} V.A.Avetisov, Yu.N.Zhuravlev, ``An evolutionary interpretation of the \p-adic ultrametric diffusion equation''
{\it Doklady Mathematics} {\bf 75} (2007) 453

\bibitem{AIMN2014} V.A.Avetisov, V.A.Ivanov, D.A.Meshkov, S.K.Nechaev
``Fractal Globules: A New Approach to Artificial Molecular Machines'' {\it Biophys.J.} {\bf 107} (2014) 2361

\bibitem{MAL2005} P.W.Messer, P.F.Arndt, M.L\"assig, ``Solvable Sequence Evolution Models and Genomic Correlations'' {\it Phys.Rev.Lett.} {\bf 94} (2005) 138103

\bibitem{DD2009} B.Dragovich, A.Yu.Dragovich ``A \p-adic model of DNA sequence and genetic code'' {\it p-Adic Numbers, Ultrametric Analysis, and Applications} {\bf 1} (2009)  34

\bibitem{GO2011}  B.Gutkin, V.Al.Osipov 
``Spectral Problem of Block-rectangular Hierarchical Matrices''
{\it J.Stat.Phys.} {\bf 143} (2011) 72

\bibitem{K2002} S.V.Kozyrev, ``Wavelet analysis as a \p-adic spectral analysis'' {\it Izvestia Akademii Nauk, Seria Math.} {\bf 66} (2002) 149

\bibitem{KKS2014} S.V.Kozyrev, A.Yu.Khrennikov, V.M.Shelkovich ``\p-Adic wavelets and their applications'' {\it Proceedings of the Steklov Institute of Mathematics} {\bf 285} (2014,) 157

\bibitem{N2014} G.Navarro ``Wavelet Trees for All'', {\it Lecture Notes in Computer Science} {\bf 7354} (2014) 2

\bibitem{St1999} R.P.Stanley, ``Enumerative combinatorics'', Vol.2, (Cambridge University Press, Cambridge, 1999)

\bibitem{W} E.W.Weisstein,  ``Line Graph'', {\it From MathWorld--A Wolfram Web Resource}  http://mathworld.wolfram.com/LineGraph.html

\bibitem{AEB1951} T.van Aardenne-Ehrenfest, N.G.de Bruijn, ``Circuits and trees in oriented linear graphs'' {\it Simon Stevin} {\bf 28} (1951) 203

\bibitem{R2002} V.R.Rosenfeld ``Enumerating de Bruijn sequences'' {\it Communications in Math. and in Computer Chem.} {\bf 45} (2002) 71

\bibitem{WXY1986} Z.Wan, R.Xiong, M.Yu
``On the number of cycles of short length in the de Bruijn-Good graph $G_n$'',
{\it Discrete Mathematics} {\bf 62} (1986) 85

\bibitem{GO2013b} B.Gutkin, V.Al.Osipov 
``Clustering of periodic orbits and ensembles of truncated unitary matrices''
{\it J. Stat. Phys.} \textbf{153} (2013) 1049;

\bibitem{NBMSHH2007} T.Nagao, P.Braun, S.M\"uller, K.Saito, S.Heusler, F.Haake, ''Semiclassical theory for parametric correlation of
energy levels'' {\it J.Phys. A: Math. Theor.} {\bf 40} (2007) 47

\bibitem{S2010}  R.Sharp ``Degeneracy in the length spectrum for metric graphs'' {\it Geometriae Dedicata Volume} {\bf 149} (2010) 177%-188

\bibitem{T2000}  G.Tanner ``Spectral statistics for unitary transfer matrices of binary graphs'' {\it J. Phys. A} {\bf 33} (2000) 3567%-3586

\bibitem{KR2000} S.Kapoor, H.Ramesh, ``An Algorithm for Enumerating All Spanning Trees of a Directed Graph'' {\it Algorithmica} {\bf 27} (2000) 120

\bibitem{H1950} R.W.Hamming ``Error Detecting and Error Correcting Codes'' {\it Bell System Technical Journal} {\bf 29} (1950) 147

\end{thebibliography}
\end{document}